\documentclass[11pt,a4paper]{article} 
\pdfoutput=1
\usepackage{jcappub}
\usepackage{subfigure}
\newcommand{\eq}[1]{\begin{equation}#1\end{equation}}
\newcommand{\eqa}[1]{\begin{eqnarray}#1\end{eqnarray}}
\newcommand{\secs}[1]{\section{#1\label{sec-#1}}}
\newcommand{\ssecs}[1]{\subsection{#1\label{ssec-#1}}}

\newcommand{\fig}[4]{\begin{figure}[#4]\centering\includegraphics[width=#3\textwidth]{#1}\caption{#2}\label{fig-#1}\end{figure}}
\newcommand{\figa}[3]{\begin{figure}[#3]\centering #1\caption{#2}\end{figure}}
\newcommand{\figi}[3]{\subfigure[#2]{\includegraphics[width=#3\textwidth]{#1}\label{fig-#1}}}

\newcommand{\refeq}[1]{eq.\ (\ref{eq-#1})}
\newcommand{\refsec}[1]{section \ref{sec-#1}}
\newcommand{\refssec}[1]{section \ref{ssec-#1}}

\newcommand{\refig}[1]{figure \ref{fig-#1}}

\newcommand{\Refig}[1]{Figure \ref{fig-#1}}

\newcommand{\subs}[1]{_\mathrm{#1}}
\newcommand{\sups}[1]{^\mathrm{#1}}
\newcommand{\dd}[1]{\mathrm{d}#1}
\newcommand{\mpl}{M\subs{p}}
\newcommand{\cL}{\mathcal{L}}
\newcommand{\hf}{\frac{1}{2}}
\newcommand{\cm}[1]{}
\newcommand{\mchi}{m_\chi}
\newcommand{\mphi}{m_\phi}
\newcommand{\nxx}[2]{N_{#1\rightarrow#2}}

\def\defaultfigsize{0.7}

\begin{document}
\title{Preheating and locked inflation: an analytic approach towards parametric resonance}
\author[a,b]{Lingfei Wang}
\affiliation[a]{Physics Department, Lancaster University, Lancaster LA1 4YB, UK}
\affiliation[b]{School of Physics, Nanjing University, 22 Hankou Road, Nanjing 210093, China}
\emailAdd{l.wang3@lancaster.ac.uk}
\abstract{We take an analytic approach towards the framework of parametric resonance and apply it on preheating and locked inflation. A two-scalar toy model is analytically solved for the $\lambda\phi^2\chi^2$ coupling for the homogenous modes. The effects of dynamic universe background and backreaction are taken into account. We show the average effect of parametric resonance to be that $\chi$'s amplitude doubles for each cycle of $\phi$.

Our framework partly solves the broad resonance for preheating scenario, showing two distinct stages of preheating and making the parameters of preheating analytically calculable. It is demonstrated for slowroll inflation models, preheating is terminated, if by backreaction, typically in the 5th e-fold. Under our framework, a possible inhomogeneity amplification effect is also found during preheating, which both may pose strong constraints on some inflationary models and may amplify tiny existing inhomogeneities to the desired scale. For demonstration, we show it rules out the backreaction end of preheating of the quadratic slowroll inflation model with mass $m\sim10^{-6}$. For locked inflation, parametric resonance is found to be inhibited if $\phi$ has more than one real component.}
\maketitle
\secs{Introduction}
Inflation was first proposed in \cite{Guth:1980zm} 30 years ago to solve multiple problems previously encountered by the Hot Big Bang theory. It soon received public attention and subsequent works of others\cite{Linde:1981mu,Linde:1983p1108,Mukhanov:1992p206,Linde:1993cn,ArmendarizPicon:1999rj,Lyth:2001nq,Linde:2001ae,Dvali:2003vv}\cm{Curvaton} have greatly perfected our understanding of inflation. Inflation has become so popular that almost all high energy theories have expressed their personal views on how inflation may be achieved, e.g. \cite{Copeland:1994vg,Stewart:1994ts,Dvali:1998pa,Kachru:2003sx,Dimopoulos:2005ac}. Even now, after 30 years of development, articles are still being entitled ``inflation'' on a daily basis.

At the same time, reheating, the immediate subsequent scenario to inflation, was more or less in oblivion. It was not until 1994 that Kofman et al.\ came to realize\cite{Kofman:1994rk} that the explosive particle production, which they named as \emph{preheating}, should be first processed before the ordinary perturbative method is applied. Three years later,  they proposed the analysis of reheating\cite{Kofman:1997yn}, covering numerous  factors taken into account. (See \cite{Allahverdi:2010xz} for a recent review.) Besides broad resonance, the narrow resonance also came into people's sight in the 1990s as a different non-perturbative particle production process during reheating\cite{Traschen:1990sw,Shtanov:1994ce}. The analytic study of preheating however remains stalled afterwards, although numerical studies and applications on inflationary models have been going on anyhow. Despite that some applications of the theory in \cite{Kofman:1997yn} have been quite successful, it is still very mathematically involved, and therefore, some of the underlying physics are not revealed.

On the other hand, locked inflation\cite{Dvali:2003vv} was proposed as a fastroll inflationary scenario. However, it was later found to be subject to multiple constraints, which together ruled out the whole parameter space\cite{Copeland:2005ey}. Among them, parametric resonance is a crucial one which may terminate locked inflation much earlier than expected. Recently, locked inflation is recalled in a cyclic universe model which incorporated D-branes and curvature\cite{Li:2011nj}, bringing locked inflation and locked inflationary contraction to each cycle. The proposed model however faces the same parametric resonance problem as locked inflation does.

In this article, we construct an analytic framework of parametric resonance for the homogenous mode, in a way more straightforward and less involved. We then apply it on preheating and locked inflation, and discuss the possible inhomogeneity amplification effect. First, from \refssec{Rolling Stage} to \refssec{Phase Delay}, we solve a half cycle of parametric resonance and show the cause of the exponential boost effect. In \refssec{Long Time Evolution}, we add up the effect cycle by cycle and give the exact solution to the Mathieu equation. We then put universe expansion into consideration in \refssec{Dynamic Background} and investigate how backreaction acts as the terminator of parametric resonance in \refssec{Backreaction}. Such approaches then allow the discussion of preheating and locked inflation in \refssec{Preheating} and \refssec{Locked Inflation} respectively. A brief discussion is carried out in \refsec{Discussion of Inhomogeneity Amplification} on how parametric resonance may amplify the existing inhomogeneity and constrain inflationary models. The results are summarized in \refsec{Summary}.

\secs{The Simple Framework}
To construct the framework, we consider the simple case first in a static background with Lagrangian density
\eq{\cL=-\hf m^2\phi^2-\hf M^2\chi^2-\lambda\phi^2\chi^2-\hf\partial^\mu\phi\partial_\mu\phi-\hf\partial^\mu\chi\partial_\mu\chi.\label{eq-L1}}
$\phi$ and $\chi$ are real scalar fields with bare masses $m$ and $M$. Their interaction strength is characterized by the dimensionless parameter $\lambda$. Here we consider the configuration of $\lambda|\phi|^2\gg m^2\gg\lambda\chi^2$ and $m^2\gg M^2$ to neglect all backreactions. $|\ |$ indicates the amplitude of oscillation when operating on a field. We allow $M^2$ to be negative, but still use $M^2$ to indicate the absolute value of $M^2$ for convenience when we compare it with other values. In many of our calculations however, we just neglect $M^2$ because it's too small. Throughout this paper, we will stick to the homogenous modes of both fields, so the spatial dependence will also be neglected in the notation. For simplicity, we will also use the word ``parametric resonance'' to indicate only the parametric resonance within our assumed parameter space. Please note this Lagrangian doesn't suffer from problems like negative infinite energy from the negative $M^2$ because it's just the effective part we take out from the total Lagrangian, so terms like the quartic ones don't distract us. As long as such terms are negligible, one doesn't need to worry about the validity of \refeq{L1}.

\figa{\figi{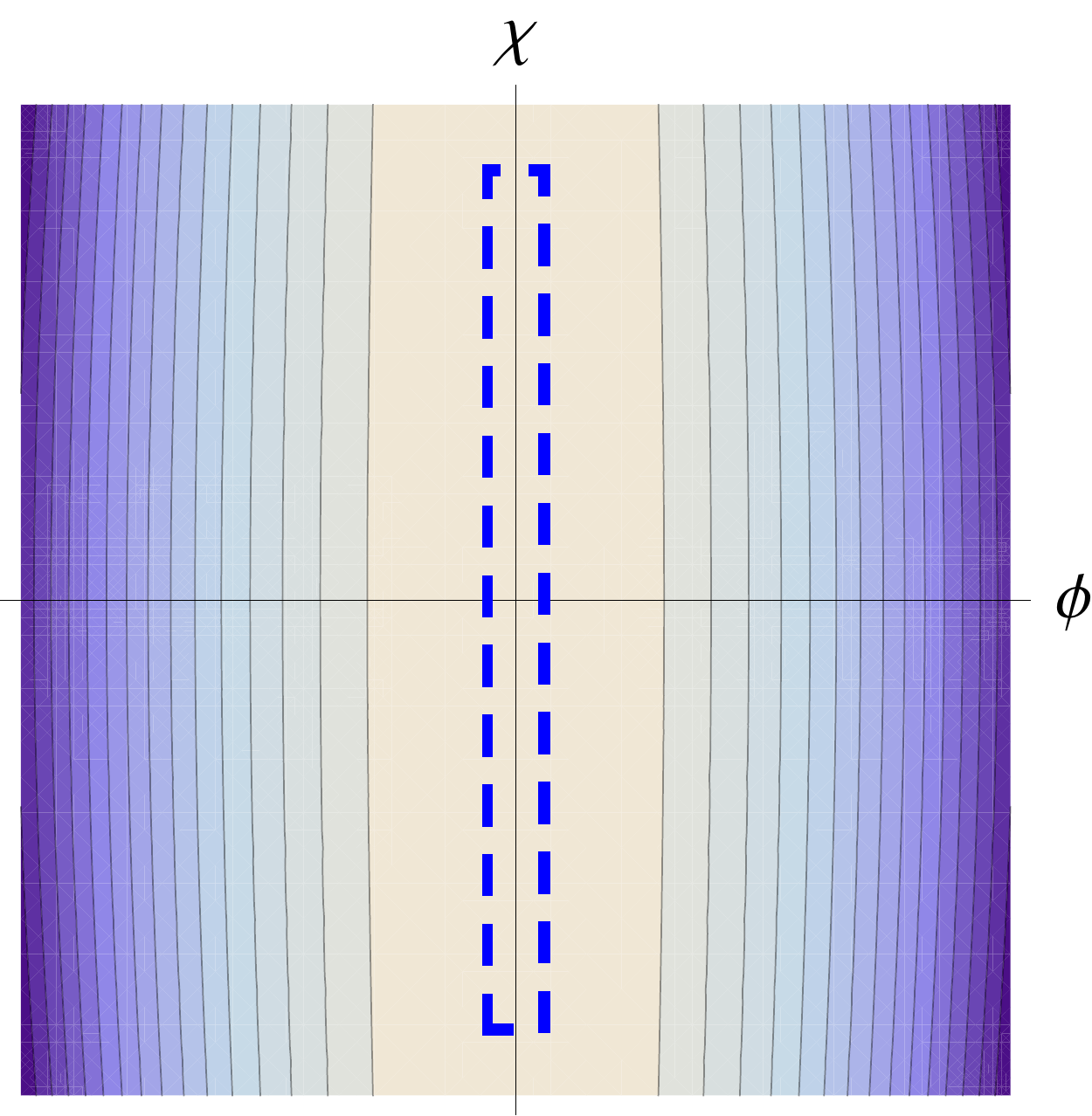}{Landscape of $V(\phi,\chi)$}{0.3}\hspace{0.04\textwidth}\figi{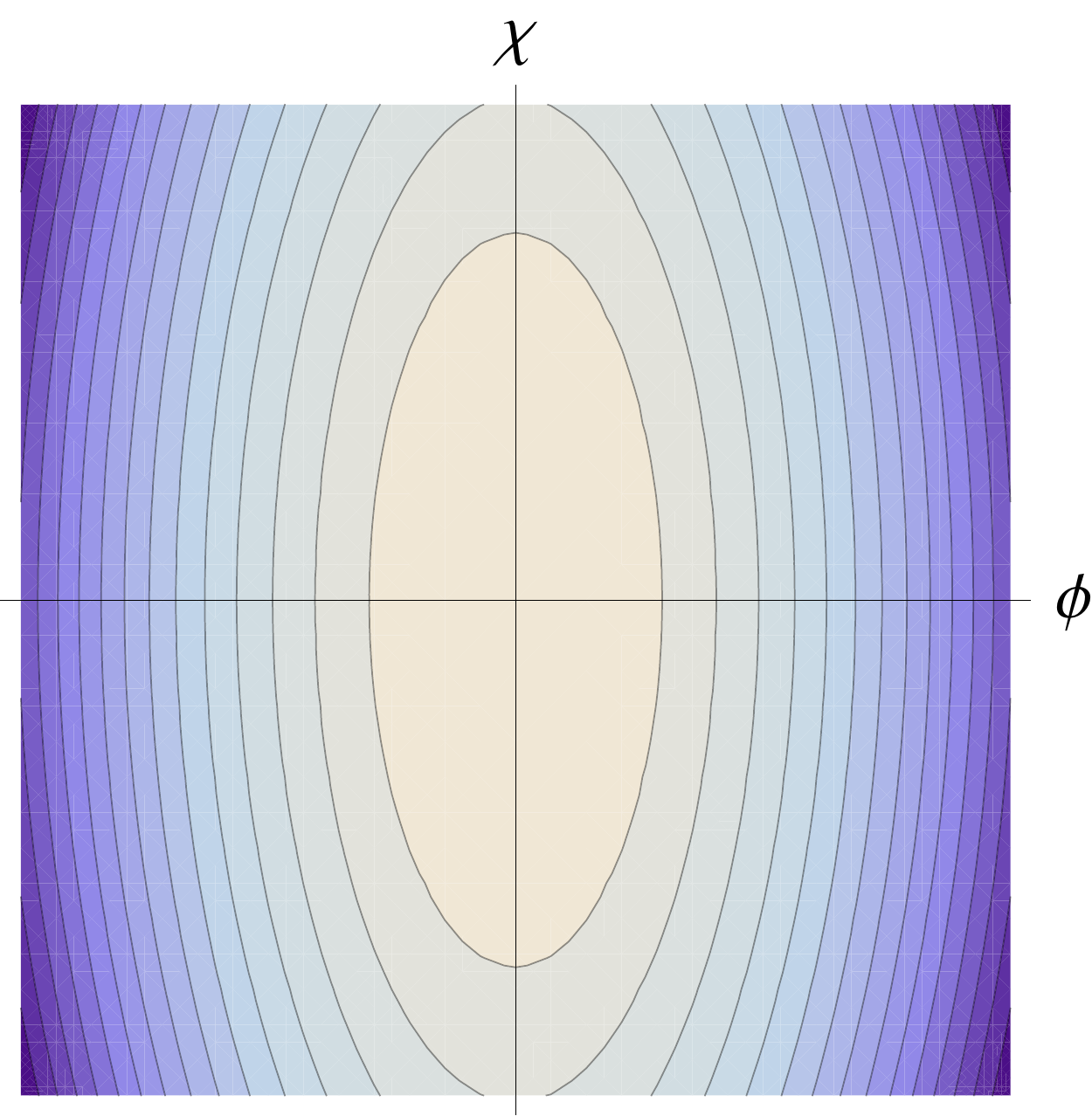}{Magnified with $M^2>0$}{0.3}\hspace{0.04\textwidth}\figi{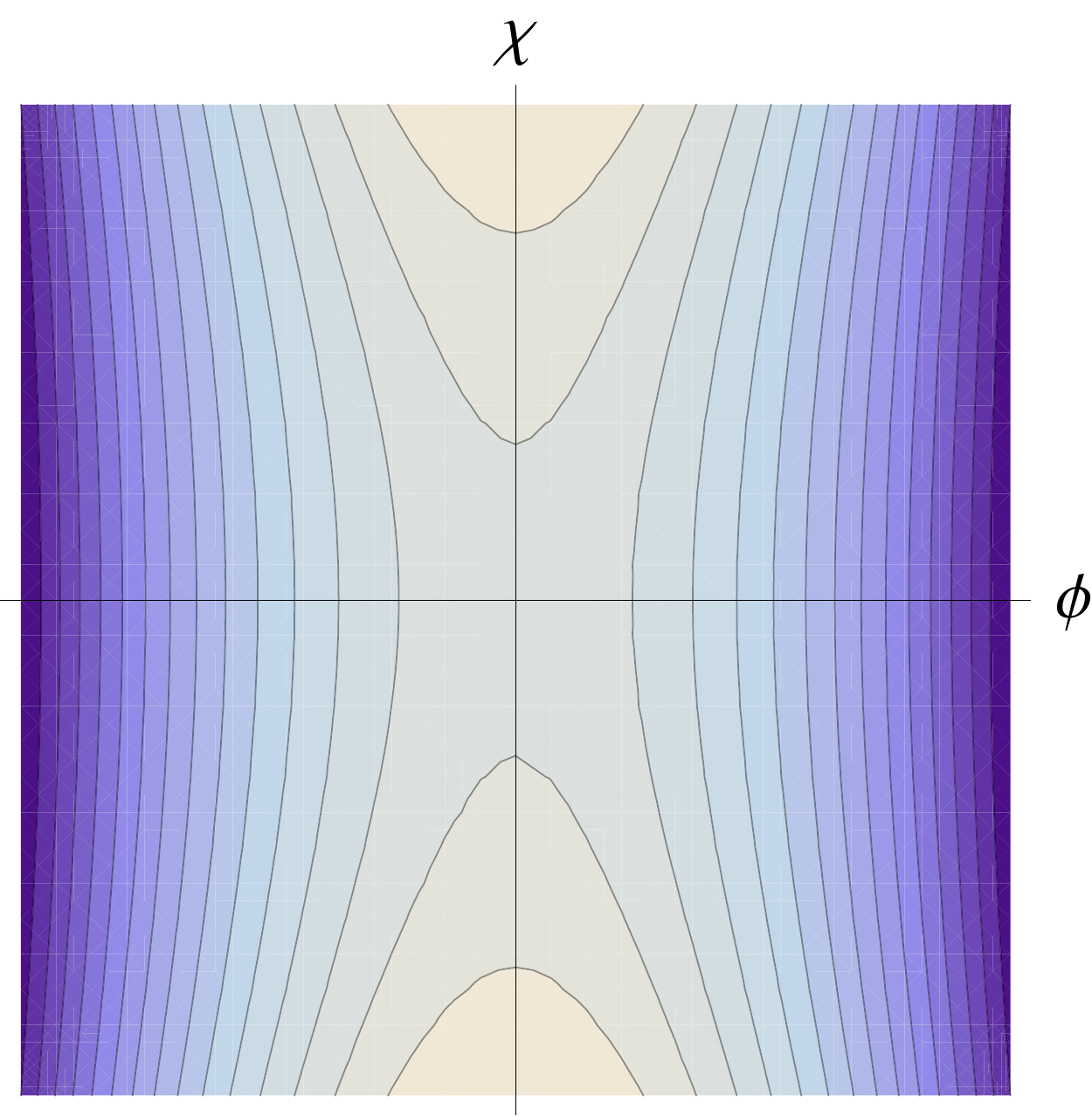}{Magnified with $M^2<0$}{0.3}}{Contour-plotting of $V(\phi,\chi)$. The first figure gives the overall landscape. The next two are zoomed in to the marked region and redrawn for details with different signs of $M^2$. A darker color implies a larger potential energy.}{t}

From the Lagrangian, we can write down the potential
\eq{V(\phi,\chi)=\hf m^2\phi^2+\hf M^2\chi^2+\lambda\phi^2\chi^2,}
in which $m^2\phi^2/2$ dominates according to the above configuration. $V(\phi,\chi)$ is visualized in \refig{Graph-PotentialLarge.pdf}, where the vertical lines also suggests the domination of $m^2\phi^2/2$. More details of the potential are displayed in \refig{Graph-PotentialSmallPositive.pdf} and \refig{Graph-PotentialSmallNegative.pdf}.

It is not difficult to imagine the motion of the two fields under such a configuration. Under $\lambda\chi^2\ll m^2$, $\phi$'s effective mass squared would be $\mphi^2\equiv m^2+2\lambda\chi^2\approx m^2$ and $\phi$ will always be oscillating with frequency $m$ and constant amplitude, unaffected by the motion of $\chi$. We can easily derive its equation of motion
\eq{\ddot\phi+(m^2+2\lambda\chi^2)\phi=0,\label{eq-eom-phi}}
where dots mean derivatives w.r.t time. After neglecting $\chi$, we get the Simple Harmonic Oscillator(SHO) solution $\phi(t)=|\phi|\sin mt$.

$\chi$'s effective mass squared $\mchi^2\equiv 2\lambda\phi^2+M^2\approx2\lambda\phi^2$ is much larger than $\phi$'s most of the time, so it will oscillate much faster than $\phi$, as long as $\phi$ is not very close to zero. When $\phi$ crosses zero however, the motion of $\chi$ would become very different from the stage of large $\phi$. We will model the two stages in turn, trying to find out why and how $\chi$ is boosted exponentially. Still, we write the equation of motion of $\chi$ here for future reference.
\eq{\ddot\chi+(M^2+2\lambda\phi^2)\chi=0.\label{eq-eom-chi}}

\ssecs{Rolling Stage}
The rolling stage is when $\phi$ is not very close to zero, therefore $\chi$ is oscillating much faster than $\phi$. It also requires that $\chi$'s effective mass varies slowly and thus can be regarded as a constant in each cycle of $\chi$'s oscillation. For most cases in which the rolling stage takes up more than half of $\phi$'s oscillation, one only needs to take into account the second condition, and then the first one is automatically satisfied. So the condition of this stage is
\eq{\left|\frac{\dd\mchi^2}{\mchi^3\dd t}\right|\ll 1.\label{eq-rs-condition}}

Although this stage can be easily solved with simple approximations, we still would like to work on it again here, mainly to arrive at the variables we are interested in and confirm the rolling stage has a generally vanishing total effect on them. Suppose at time $t=0$, $\chi$ is at maximum with $\chi=|\chi|$ and $\dot\chi=0$, and scalar field $\phi$ is rolling down slowly with time derivative $\dot\phi$. We take the effective mass $\mchi$ constant and get the zeroth order SHO solution of \refeq{eom-chi}
\eq{\chi\sups{(0)}\equiv|\chi|\cos \mchi t.}
We then derive the equation of motion up to first order $\chi\sups{(1)}$ by introducing $\dot\phi$,
\eq{\ddot\chi\sups{(1)}+\mchi^2\chi\sups{(1)}+4\lambda\phi\dot\phi|\chi|t\cos \mchi t=0}
and get the solution with the initial condition $\chi\sups{(1)}(0)=0$ and $\dot\chi\sups{(1)}(0)=0$,
\eq{\chi\sups{(1)}(t)=-\frac{\lambda\phi\dot\phi|\chi|}{\mchi^3}\biggl(\mchi t\cos \mchi t+(\mchi^2t^2-1)\sin \mchi t\biggr).}
At time $t=\pi/\mchi$, when $\chi\sups{(0)}$ reaches maximum again, we have
\eq{\chi\sups{(1)}\Bigl(\frac{\pi}{\mchi}\Bigr)=\frac{\pi\lambda\phi\dot\phi|\chi|}{\mchi^3},\hspace{0.5in}\dot\chi\sups{(1)}\Bigl(\frac{\pi}{\mchi}\Bigr)=\frac{\pi^2\lambda\phi\dot\phi|\chi|}{\mchi^2}.}

If we take when $\phi$ is rolling down as an example, we will find out $\chi$ gets an additional amplitude correction from the decrease of $\phi$. To be precise, we consider the time interval $\Delta t\equiv \pi/\mchi$, during which $\chi$ rolls from one maximum to the other. $\chi$'s amplitude is affected at a rate
\eq{\frac{\Delta|\chi|}{\Delta t}=-\frac{\chi\sups{(1)}(\frac{\pi}{\mchi})}{\Delta t}=-\frac{\lambda\phi\dot\phi}{\mchi^2}|\chi|.}
Replacing $\Delta$ with differential operator gives
\eq{|\chi|\propto \frac{1}{\sqrt\mchi},\label{eq-rs-chi-propto-mass}}
which is the effect on $|\chi|$ due to the variation of $\phi$.

Similarly, we can calculate $\chi$'s phase change resulting from the variation of $\phi$, or $\mchi$ directly. The motion of $\phi$ generates an additional $\dot\chi\sups{(1)}(\pi/\mchi)$ which changes the time needed to push $\chi$ back to its maximum again. The phase change is therefore
\eq{\varphi_\chi\equiv\frac{\dot\chi\sups{(1)}(\frac{\pi}{\mchi})}{\mchi|\chi|}=\frac{\pi^2\lambda\phi\dot\phi}{\mchi^3}=\frac{\pi}{2}\log\frac{\mchi'}{\mchi},\label{eq-rs-chi-phase-effect}}
in which $\mchi$ and $\mchi'$ are the initial and final effective masses of $\chi$.

The above calculation has shown how $\chi$ is affected by $\phi$'s motion when $\phi$ is away from zero. Although the effects on $\chi$'s amplitude (\refeq{rs-chi-propto-mass}) and phase (\refeq{rs-chi-phase-effect}) are strong, $\chi$ can't get exponentially boosted in this way. This is because the effects during the upward rolling period of $\phi$ are exactly inverse from those during the downward period, so they generally cancel. $\chi$ is increased when $\phi$ is rolling towards zero, but it get decreased when $\phi$ moves away from zero. The same thing happens to $\chi$'s phase change, so the net effect vanishes.

\ssecs{Zero-Crossing Stage}
When $\phi$ is small and $\dot\phi$ is relatively large, the condition of rolling stage \refeq{rs-condition} breaks, and the system enters the zero-crossing stage. The zero-crossing stage is the period $\phi$ crosses zero. During this stage, the motion of $\chi$ is very much different from that during the rolling stage. It's because $\chi$'s effective mass $\mchi$ varies very fast in this stage and can become smaller than $\phi$'s effective mass $\mphi$ when $\phi$ is crossing zero.

Because $m^2\gg\lambda\chi^2$, we still adopt the approximation that $\phi$ is not affected by the motion of $\chi$, and $M^2\ll \lambda|\phi|^2$ is still negligible. Here we set $t=0$ at $\phi=0$. In most applications of parametric resonance, $|mt|\ll 1$ holds throughout the zero-crossing stage. We then get the motion of the two fields
\eq{\phi=|\phi|mt,}
and thus
\eq{\ddot\chi+2\lambda|\phi|^2m^2t^2\chi=0.}
Here we use $\tau\equiv t(2\lambda|\phi|^2m^2)^\frac{1}{4}$ as the scaled time, and primes as derivatives w.r.t $\tau$. The equation of motion of $\chi$ then becomes
\eq{\chi''+\tau^2\chi=0.\label{eq-zcn-eomchi}}

The breaking of condition \refeq{rs-condition} corresponds to $|\tau|\sim<1$. If one is only interested in the behavior of the system in $|\tau|<1$, one can simply expand $\chi$ in Taylor series of $\tau$ at $\tau=0$ and only take the first few terms. Numerical simulation tells us, however, the energy boost on $\chi$ spans not only $|\tau|<1$, but at an even broader range. To precisely calculate the boost effect, we need to take the analytic solution of \refeq{zcn-eomchi}
\eq{\chi(\tau)={}_0F_1\Bigl(\frac{3}{4},-\frac{1}{16}\tau^4\Bigr)\chi_0+{}_0F_1\Bigl(\frac{5}{4},-\frac{1}{16}\tau^4\Bigr)\chi_0'\tau,\label{eq-zcn-chi}}
where $_0F_1(a,z)$ is the \emph{confluent hypergeometric (limit) function}, and the subscript 0 denotes the value of variables at $\tau=0$.

Given the initial condition $\chi_0$ and $\chi_0'$, one can already calculate the effect of parametric resonance. In order to enable our further calculation, we take out the leading term of $\chi$ for $|\tau|\gg 1$ (but $|mt|\ll1$ still holds). To do this, we expand $_0F_1(a,-z)$ in series of $1/z$. The leading term is
\eq{{}_0F_1(a,-z)\approx\frac{\Gamma(a)}{\sqrt\pi}\sin\Bigl(\frac{\pi}{4}(2a+1)-2\sqrt z\Bigr)z^{\frac{1}{4}-\frac{a}{2}},\hspace{0.3in}z\gg1.\label{eq-zcn-0F1}}

Here we want to calculate the energy boost effect from parametric resonance, so we characterize the boost rate with
\eq{\eta\equiv\hf\lim_{\tau\rightarrow+\infty}\log\frac{E_s(\tau)}{E_s(-\tau)}.\label{eq-zcn-etadef}}
The scaled energy density of $\chi$ is defined as
\eq{E_s(\tau)\equiv \tau^2\chi^2(\tau)+\chi'^2(\tau),}
with its relation with the (unscaled) total energy density of $\chi$
\eq{E\equiv\lambda|\phi|^2m^2t^2\chi^2+\hf\dot\chi^2=\sqrt\frac{\lambda|\phi|^2m^2}{2}E_s.}
Substituting \refeq{zcn-chi} and \refeq{zcn-0F1} into \refeq{zcn-etadef}, we arrive at
\eq{\eta=\hf\log\biggl(1+\frac{4\sqrt2\Gamma(\frac{3}{4})\Gamma(\frac{5}{4})\chi_0\chi_0'}{\Gamma^2(\frac{3}{4})\chi_0^2-2\sqrt2\Gamma(\frac{3}{4})\Gamma(\frac{5}{4})\chi_0\chi_0'+4\Gamma^2(\frac{5}{4})\chi_0'^2}\biggr).\label{eq-zcn-eta}}

In the above equation, the limit $\tau\rightarrow+\infty$ has already been taken. As a result, all non-leading terms of $\tau$ are removed. The $2\sqrt z$ in the sin function of ${}_0F_1$ approximation is  removed too.

From \refeq{zcn-eta} we find the boost rate is hardly dependent on any parameter of the system, ($m$, $\lambda$, etc,) or the energy density of any component, as long as our previous assumptions stay valid. It is also easily verified by numerical simulations. This constant property is very helpful when we sum the boost rates across cycles of $\phi$ in subsequent calculations. The fraction in the log function of \refeq{zcn-eta} should be of order unity, so the boost effect is significant and exponential. The boost rate $\eta$ acquires the same sign with $\chi_0\chi_0'$, which means the energy of $\chi$ can either be boosted or dampened, depending on the sign of $\chi_0\chi_0'$.

\ssecs{Phase Delay}
In the above subsection, we have represented $\eta$ with $\chi_0$ and $\chi_0'$, the motion state of $\chi$ at $\phi=0$. The previous calculation however only gives half of the analytic solution --- we can calculate $\eta$ with any given $\chi_0$ and $\chi_0'$, but we haven't developed any method in deriving them two, except numerical ways. Therefore we will derive $\chi_0$ and $\chi_0'$ analytically in this subsection and provide a completely analytic solution. But before that, it's quite important to first clarify what ``phase'' is referred to here.

The system has two degrees of freedom, $\phi$ and $\chi$, which means we need four independent parameters to fully describe the motion state of the system at any specific time. We are interested in the two parameters characterizing $\chi$, which can either be chosen as $\chi$ and $\chi'$, or its amplitude and phase. From \refeq{zcn-eta}, we have already learned the boost rate $\eta$ is a function of $\chi_0$ and $\chi_0'$. Although $\chi$ is not strictly a SHO at the zero-crossing stage, if we manage to represent the motion of $\chi$ at $\phi=0$ with an amplitude-like and a phase-like variable however, the amplitude-like variable would cancel and $\eta$ would then only depend on the phase-like variable. It's obvious $\eta$ may not stay constant in every half cycle of $\phi$, so reducing $\eta$ to a single-parameter function allows us to track $\eta$ better when studying the boost effect across many cycles of $\phi$.

In order to find the phase representation of $\chi$ at $\phi=0$, we attempt to solve $\chi_0$ and $\chi_0'$ reversely. It is reasonable to believe $\chi$ is SHO-like at $|\tau|\gg1$ and set the initial condition as
\eq{\chi(\tau)=\frac{\sqrt{E_s(\tau)}}{|\tau|}\sin\beta,\hspace{0.5in}\chi'(\tau)=\sqrt{E_s(\tau)}\cos\beta\label{eq-pdn-initcond}}
where $\beta$ is the variable we use to represent the phase of $\chi$ at $\tau$. We are only interested in $\beta$ here, so we come up with the combination
\eq{|\tau|\chi\cos\beta-\chi'\sin\beta=0.}
By putting in the solution \refeq{zcn-chi} and \refeq{zcn-0F1}, it simplies to
\eq{\Gamma\Bigl(\frac{3}{4}\Bigr)\sin\Bigl(\beta+\hf \tau^2-\frac{5}{8}\pi\Bigr)\chi_0=2\Gamma\Bigl(\frac{5}{4}\Bigr)\sin\Bigl(\beta+\hf \tau^2-\frac{7}{8}\pi\Bigr)\chi_0'.\label{eq-pdn-betaeq}}

For a specific process with initial conditions given, $\chi_0$ and $\chi_0'$ should be definite and thus independent of the choice of $\tau$ in \refeq{pdn-betaeq}. For this reason, we contract the $\tau^2/2$ into $\beta$ and redefine $\beta$ as
\eq{\beta=\beta(\tau)\equiv\beta_0-\hf \tau^2+\frac{7}{8}\pi,}
where $\beta_0$ is the variable for phase but it's now independent of time $\tau$. For short, we will however still use $\beta$ to indicate $\beta_0$. In this way, \refeq{pdn-betaeq} also becomes independent of $\tau$, and simplifies to
\eq{\chi_0\Gamma\Bigl(\frac{3}{4}\Bigr)\cos\Bigl(\beta-\frac{\pi}{4}\Bigr)=2\chi_0'\Gamma\Bigl(\frac{5}{4}\Bigr)\sin\beta.}
This immediately gives the solution
\eqa{
\chi_0&=&2\Gamma\Bigl(\frac{5}{4}\Bigr)\sin\beta\,X,\label{eq-pdn-chi0}\\
\chi_0'&=&\Gamma\Bigl(\frac{3}{4}\Bigr)\cos\Bigl(\beta-\frac{\pi}{4}\Bigr)\,X,\label{eq-pdn-chi0p}}
where $X$ is a function of $\beta$ and $E_s(\tau)$ and is not important in our calculation.

\fig{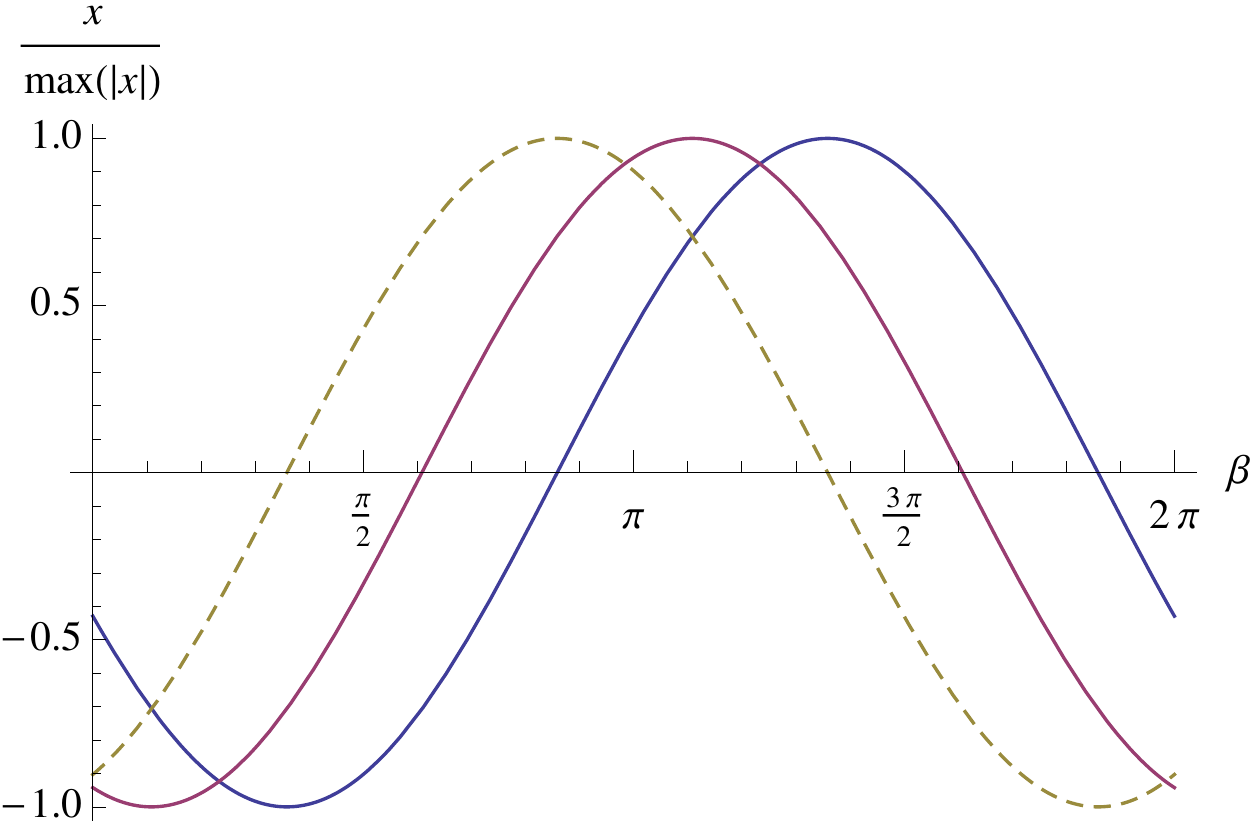}{$\chi_0$ and $\chi_0'$ as functions of $\beta$. $x$ (in the label of $y$-axis) represents $\chi_0$ and $\chi_0'$ for the blue and purple curve respectively. The max goes over all $\beta\in[0,2\pi)$ so all functions are normalized to be maximally at unity. The dashed curve is the expected $\chi_0'$ curve derived from $\chi_0$ if $\chi$ were a SHO all the time. The purple curve's deviation from the dashed curve shows the phase delay in $\chi$'s kinetic part at $\tau=0$. Parameters are chosen as $\lambda=1,\ m=1,\ M=10^{-3},\ |\phi|=100$, and they are chosen as such by default in the following figures.}{\defaultfigsize}{t}
By comparing it with the SHO solution, or the SHO-like solution \refeq{pdn-initcond}, one can find out the motion of $\chi$ at $\tau=0$ is indeed non-SHO. The phase of its kinetic term is delayed by $\pi/4$. This is also confirmed numerically for all $\beta\in[0,2\pi)$ in \refig{Graph-PhaseThetas.pdf}. Such a phase delay can be understood from the decreasing transfer efficiency from $\chi$'s potential energy to kinetic energy as $\phi$ is approaching zero. Given the $\beta$ that makes $\chi_0=0$, i.e. $\chi$ at the bottom of its potential at $\tau=0$, such $\beta$ doesn't induce the largest $\chi_0'$ any more. To understand it better, we compare such $\beta$ with $\tilde\beta=\beta+\delta\beta$, where $0<\delta\beta\ll1$. The difference caused by $\delta\beta$ is two-fold. On one hand, the phase $\tilde\beta$ makes $\chi$ reach every value slightly earlier than phase $\beta$, and such difference in timing gives $\tilde\beta$ a larger effective mass of $\chi$ when compare at the same value of $\chi$, and in consequence, a larger energy transfer efficiency of order $\delta\beta$. On the other hand, for $\tilde\beta$, $\chi$ gets an extra up-climbing period just before $\tau=0$. This effect decreases the energy transfer efficiency of $\tilde\beta$, but only has order $\delta\beta^2$. As a result, the net effect is $\tilde\beta$ has a larger energy transfer efficiency and therefore a larger kinetic energy. The maximum of kinetic energy is then shifted towards a larger $\beta$, and its phase gets delayed relatively.

We continue our calculation by substituting the set of solution \refeq{pdn-chi0} and \refeq{pdn-chi0p} into the expression of $\eta$, \refeq{zcn-eta}, and it becomes a very simple function of $\beta$
\eq{\eta(\beta)=\hf\log\biggl(1+4\sqrt 2\sin\beta\cos\Bigl(\beta-\frac{\pi}{4}\Bigr)\biggr).\label{eq-pdn-eta}}

\fig{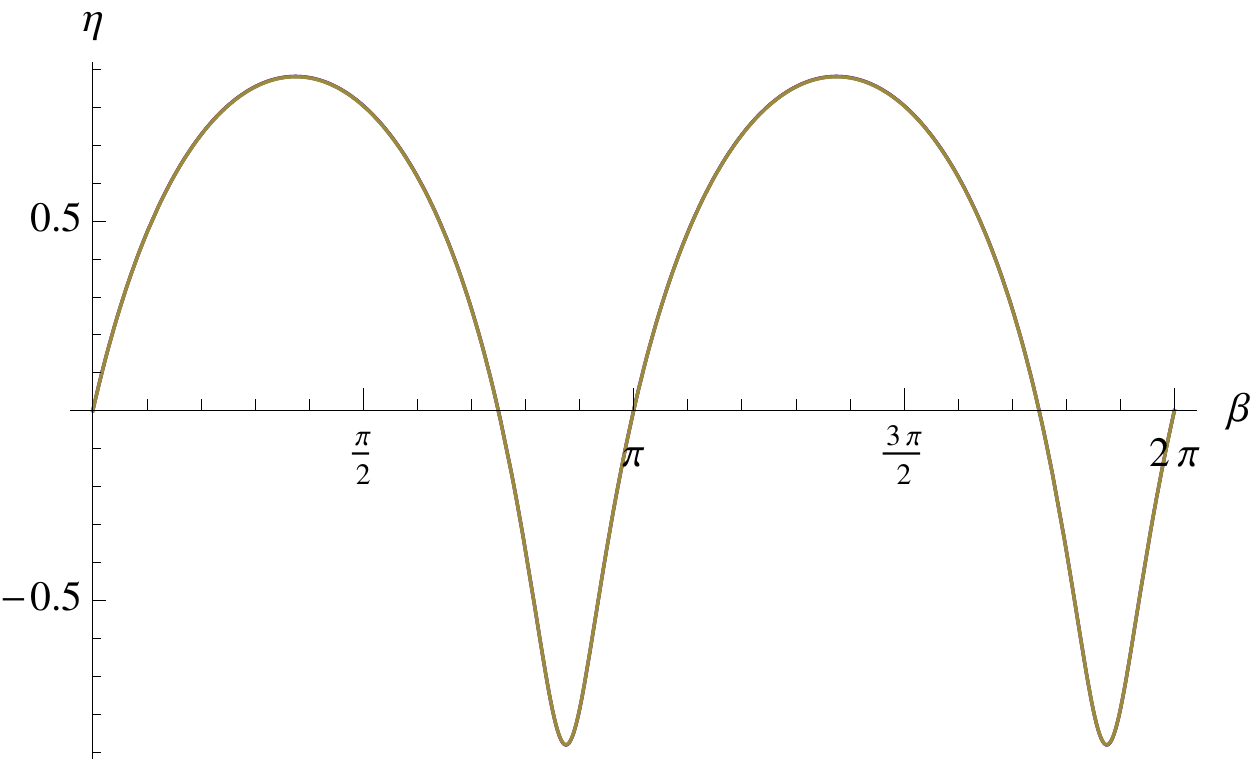}{The energy boost effect w.r.t $\beta$. All curves are horizontally shifted to be starting at $\eta=0$ and $\dd\eta/\dd\beta>0$ when $\beta=0$, so they are aligned. The curves of analytical solution, numerical simulation and semi-analytic result overlap perfectly so only one curve is visible.}{\defaultfigsize}{t}
We are now able to compare our results with numerical ones. In \refig{Graph-PhaseBoostModel.pdf}, three curves of $\eta$ are drawn --- the analytical result \refeq{pdn-eta}, the numerical solution, and the semi-analytical result which is acquired by substituting the numerically derived $\chi_0$ and $\chi_0'$ into \refeq{zcn-eta}. All three curves match perfectly.

So now we have successfully reduced the parametric resonance effect to a completely constant one --- it depends only on $\beta$, the phase of $\chi$. Ths boost rate is however independent of any of the parameters of the system, $\lambda$, $m$, etc, or even the energy density of $\chi$. This is indeed the case from numerical simulations. Because the boost rate is independent of the energy of $\chi$, we can simply sum the boost rates when calculating the total boost rate for many cycles of $\phi$.

If we naively assume $\beta$ is evenly distributed in $[0,2\pi)$, and observe many enough independent half cycles of parametric resonance, the average boost rate can then be calculated, which is
\eq{\langle\eta\rangle\equiv\frac{1}{2\pi}\int_0^{2\pi}\eta(\beta)\dd\beta=0.346.\label{eq-pdn-etaa}}
This actually means for every half cycle of $\phi$, $\chi$'s energy density becomes $e^{2\langle\eta\rangle}=2$ times of its original. For future reference, here we also define the maximum boost rate
\eq{\eta_m\equiv\max_{\beta\in[0,2\pi)}\eta(\beta)=\eta\Bigl(\frac{3}{8}\pi\Bigr)=\log(1+\sqrt2)=0.881.\label{eq-pdn-etam}}

\ssecs{Long Time Evolution}
We now continue to study the long time evolution of $\chi$ across cycles of $\phi$. Suppose the initial phase of $\chi$ is $\beta$, by which we mean initially, when $\phi$ reaches its maximum, there is
\eq{\chi=\sqrt\frac{E}{\lambda|\phi|^2}\sin\beta,\hspace{0.5in}\dot\chi=\sqrt{2E}\cos\beta.}
We can then use the boost rate $\eta(\beta)$ acquired in the last subsection to evolve the system from one maximum of $\phi$ to the next. To characterize the state of $\chi$, at every maximum of $\phi$ we use a polar coordinate system as the phase diagram of $\chi$. The angular coordinate is chosen to be $\beta$, and the radial coordinate is $e^{\eta_a}$, where $\eta_a\equiv\sum\limits_{previous}\eta$ is the total boost rate. Therefore the $x$ and $y$ coordinates are proportional to $\dot\chi$ and $\chi$ respectively. If we choose the initial condition as a constant energy density with evenly distributed $\beta$, it then corresponds to a smooth circle with unit radius on the phase diagram. For a general overview, numerical results are given in \refig{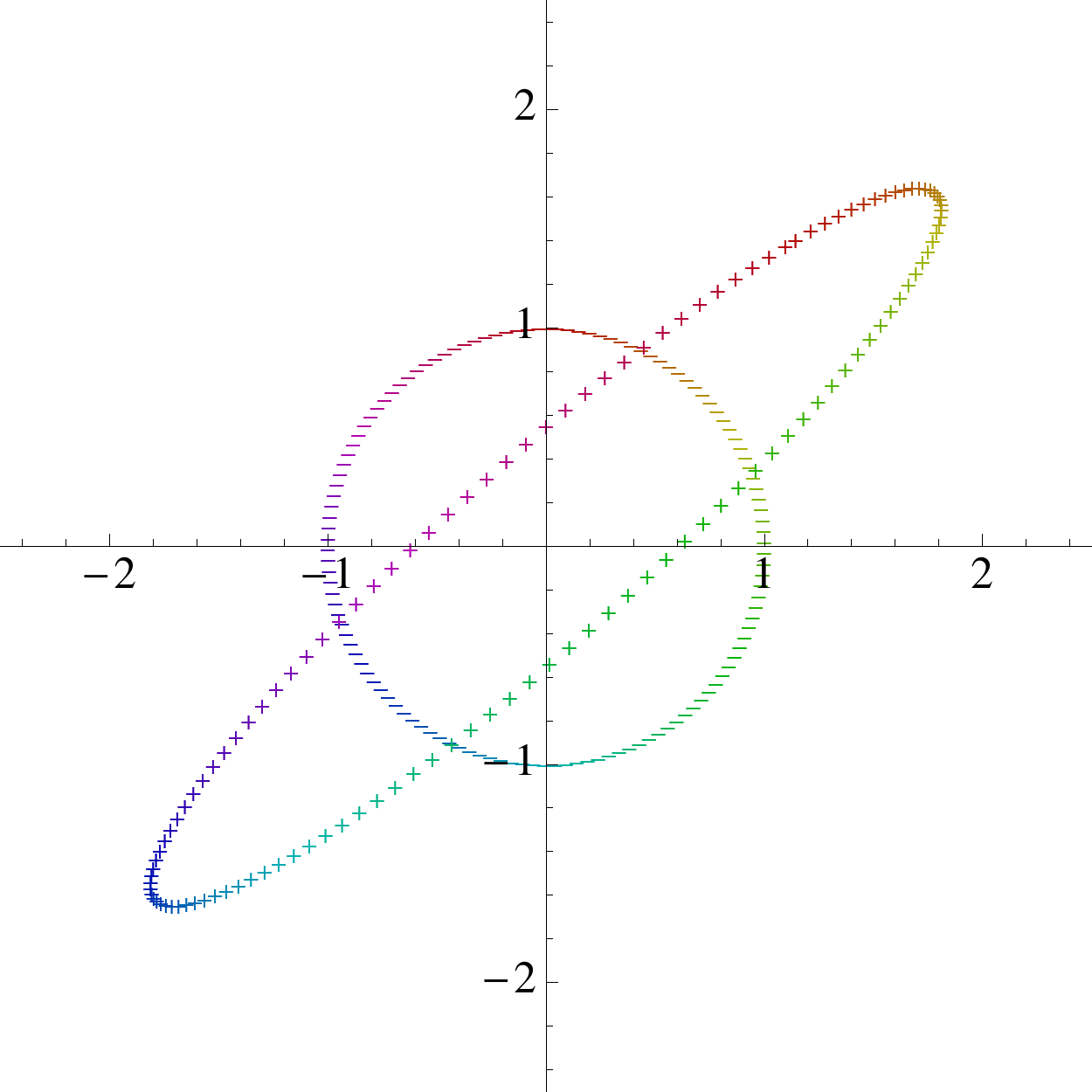}, \refig{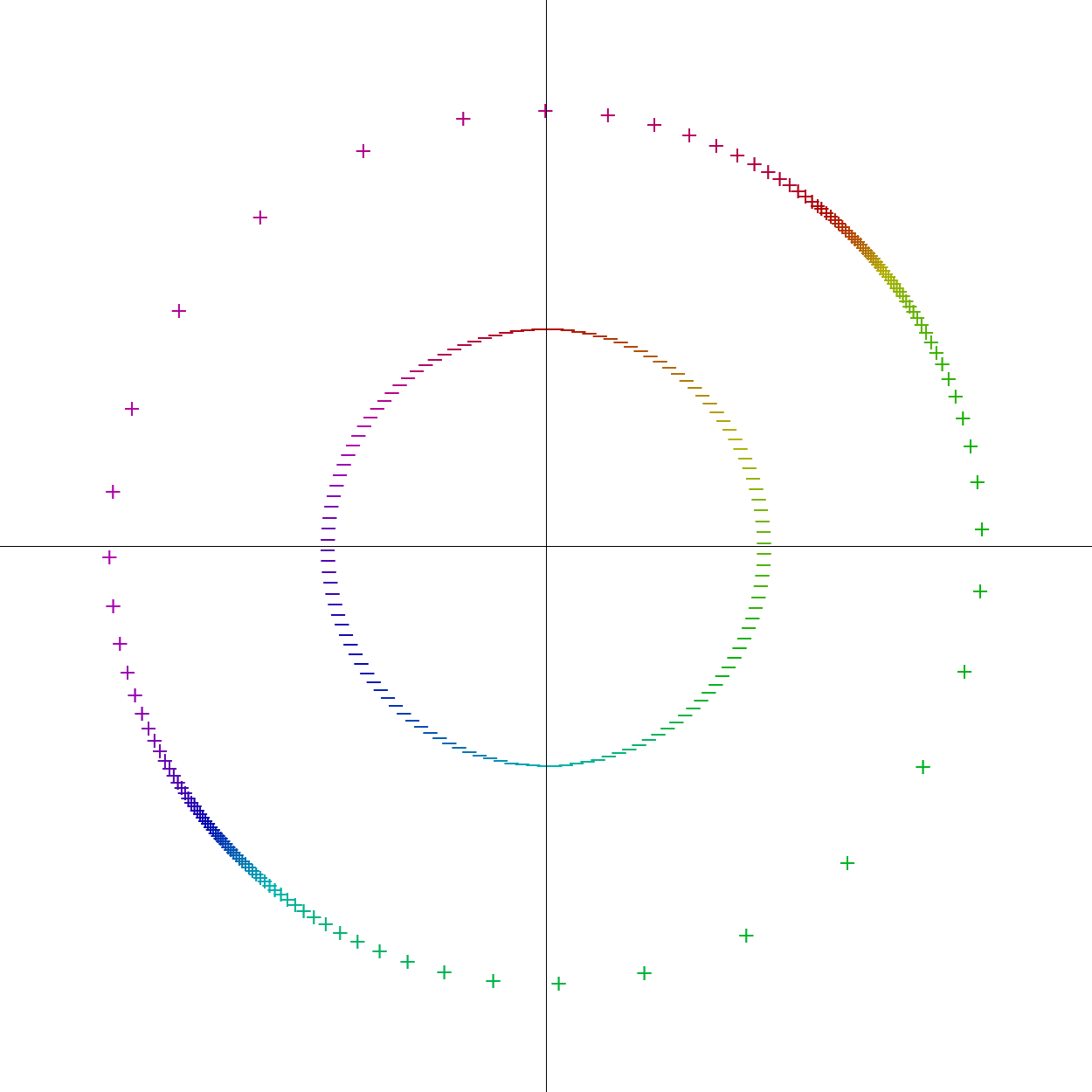} and \refig{Graph-PhaseChange2}.

\fig{Graph-PhaseChangeBoth1.pdf}{Phase diagram of $\chi$ at the neighboring maxima of $\phi$. The initial condition at the earlier maximum of $\phi$ is set to a smooth unit circle with mark ``--''. Sign ``+'' marks the later maximum, i.e. after half a cycle of $\phi$. Plots are colored according to their initial phase $\beta$, so every ``--'' mark evolves to its corresponding ``+'' mark with the same color. The rotation of colors indicates the phase shift --- the difference between the $\beta$'s at the neighboring maxima of $\phi$.}{\defaultfigsize}{t}
\fig{Graph-PhaseChangePhase1.pdf}{Phase redistribution over half a cycle of $\phi$. This figure aims to show how the evolution over half a cycle of $\phi$ affects the distribution of $\beta$, i.e. the phase of $\chi$. It is based on \refig{Graph-PhaseChangeBoth1.pdf}, and derived by scaling every ``+" plot to the same radius while keeping its phase $\beta$ fixed.}{0.7}{t}
\figa{
\figi{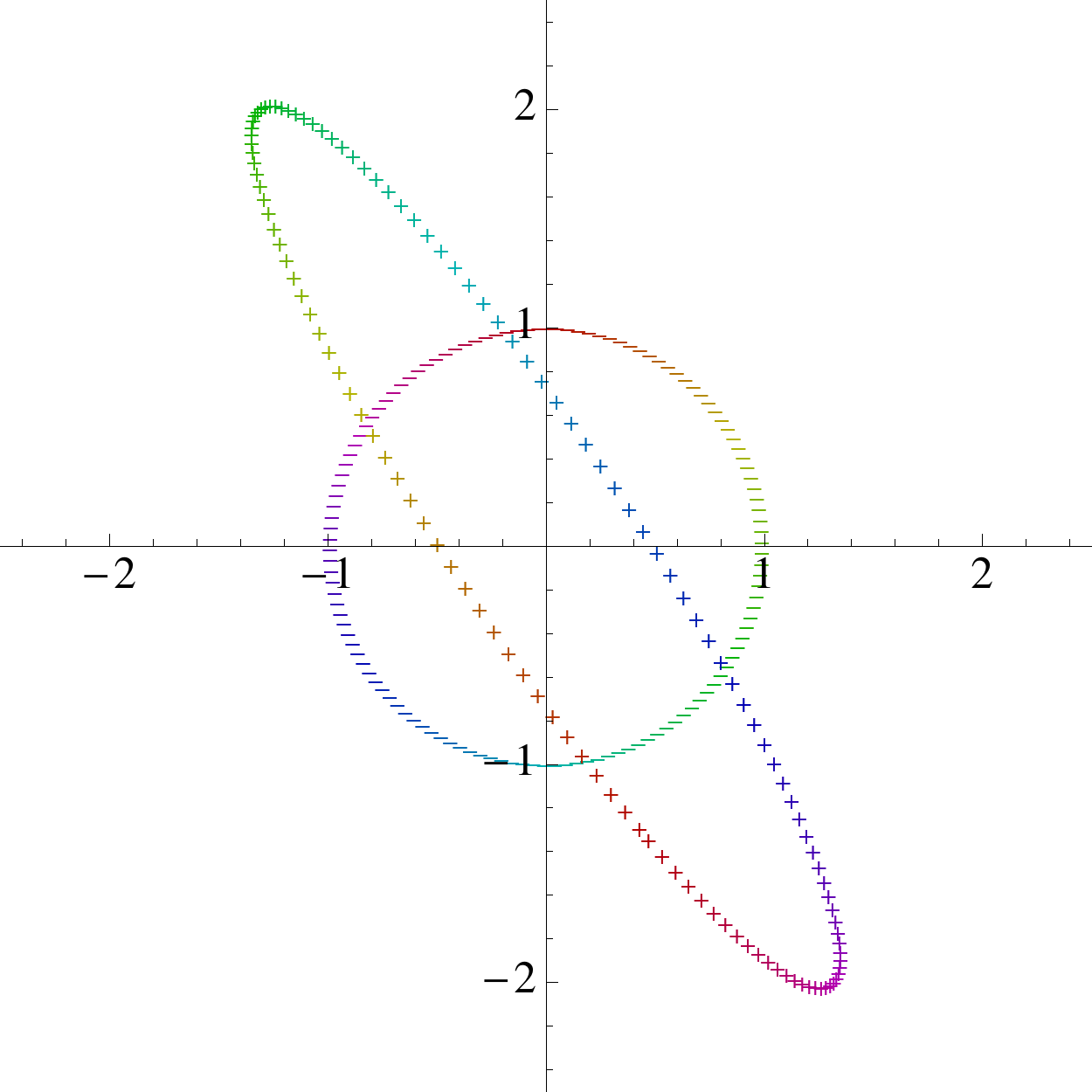}{}{0.5}\hspace{0.2in}
\figi{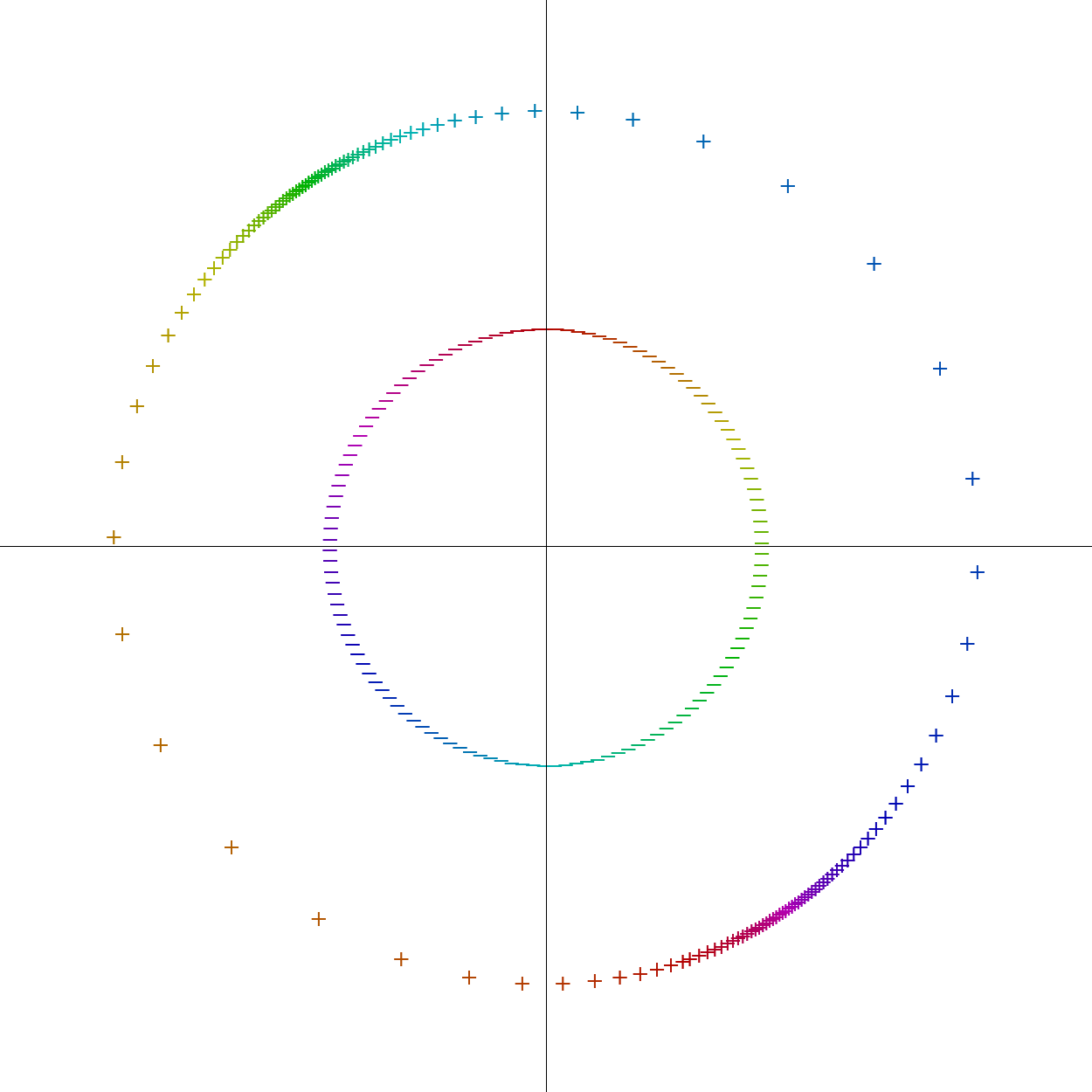}{}{0.454}
}{Redraw \refig{Graph-PhaseChangeBoth1.pdf} and \refig{Graph-PhaseChangePhase1.pdf} with $|\phi|=99$ instead of 100.\label{fig-Graph-PhaseChange2}}{t}

These figures suggest a set of linear operations be applied on each ``--'' mark on the phase diagram, in order to get to its corresponding ``+'' mark with analytical methods. The linearity is also determined by the linear equation of motion of $\chi$. To clearly demonstrate the operations are linear, we represent plots on the phase diagram with $(1,2)$ matrices $(x,y)^T$ of their Cartesian coordinates, and model the total evolution of field $\chi$ between neighboring maxima of $\phi$ as a $(2,2)$ linear transformation matrix, which we call here as the \emph{transformation matrix}. The transformation matrix should then be decomposed into two parts:
\begin{enumerate}
\item During the zero-crossing stage, $\chi$ is boosted with the boost rate $\eta$ a function of phase $\beta$. So for an initial circle on the phase diagram, it is distorted to an ellipse because of the phase dependence of $\eta$. For convenience, we rotate the ellipse by shifting $\beta$ to align its major axis with $x$-axis. If the initial circle has a radius $e^{\eta_a}$, we know at once the semi-major axis has length $e^{\eta_a+\eta_m}$, and the length of semi-minor axis is $e^{\eta_a-\eta_m}$. The largest negative boost rate $-\eta_m$ comes from the time-reversibility of $\chi$'s equation of motion. When time-reversed, the largest (positive) boost rate $\eta_m$ always corresponds to the largest negative boost rate $-\eta_m$ and proves its existence. During this step, the plot in the phase diagram is transformed to $(x',y')^T$ which satisfies
\eq{\left(\begin{array}{c}x'\\y'\end{array}\right)=\left(\begin{array}{cc}e^{\eta_m}&0\\0&e^{-\eta_m}\end{array}\right)\left(\begin{array}{c}x\\y\end{array}\right).}
\item The ellipse is rotated with an angle which is generated from the difference of the periods of $\phi$ and $\chi$. This is because the period of $\phi$ is not an integer times of the period of $\chi$. Therefore $\chi$ doesn't return to the exactly same position when $\phi$ reaches maximum (or zero) again as where it was at $\phi$'s last maximum (or zero). One can easily confirm the rotation angle is independent of phase $\beta$, so the elliptical shape is preserved. Here we suppose the rotation angle that should be added to $\beta$ is $\delta$, so the plot is transformed to
\eq{\left(\begin{array}{c}x''\\y''\end{array}\right)=\left(\begin{array}{cc}\cos\delta&-\sin\delta\\\sin\delta&\cos\delta\end{array}\right)\left(\begin{array}{c}x'\\y'\end{array}\right).}
\end{enumerate}
By definition, $\delta$ is the phase $\chi$ goes through from one maximum of $\phi$ to the next. So we can simply calculate the rotation angle $\delta$ as
\eq{\delta\equiv2\int_0^\frac{\pi}{2m}\sqrt{2\lambda}|\phi|\sin mt\dd t=\frac{2\sqrt{2\lambda}|\phi|}{m},\label{eq-pr-delta}}
in which the bare mass $M$ has been neglected. Therefore the total transformation matrix is
\eqa{\left(\begin{array}{cc}\cos\delta&-\sin\delta\\\sin\delta&\cos\delta\end{array}\right)\left(\begin{array}{cc}e^{\eta_m}&0\\0&e^{-\eta_m}\end{array}\right)
&=&\left(\begin{array}{cc}e^{\eta_m}\cos\delta&-e^{-\eta_m}\sin\delta\\e^{\eta_m}\sin\delta&e^{-\eta_m}\cos\delta\end{array}\right)\nonumber\\
&=&B^{-1}\left(\begin{array}{cc}b-\sqrt{b^2-1}&0\\0&b+\sqrt{b^2-1}\end{array}\right)B,\label{eq-pr-transmatrix}}
where $b\equiv\cosh\eta_m\cos\delta=\sqrt2\cos\delta$ and $B$ is a $(2,2)$ matrix of eigenvectors which is not important.\footnote{To be more precise, one may want to calculate the transformation matrix via a three-step method --- two rotations with angle $\delta/2$ separated by a boost. The difference however only shows in the eigenvectors $B$, but the eigenvalues remain the same.}

We are mostly interested in the total boost rate during many cycles of $\phi$, which is represented by $\eta_a=\log\sqrt{x^2+y^2}=\log\sqrt{(x,y)(x,y)^T}$. Since $(x,y)^T$ is multiplied on the left by the transformation matrix for every half cycle of $\phi$, the total boost rate is actually determined by many powers of the transformation matrix. For this purpose, we have derived \refeq{pr-transmatrix} which allows us to only consider the power of the eigenvalues of the transformation matrix.

For $b^2>1$ i.e. $\cos^2\delta>1/\cosh^2\eta_m$, we have $b-\sqrt{b^2-1}<1<b+\sqrt{b^2-1}$. Because one eigenvalue is larger than the other, the overall effect is the system converges to the eigenstate with the larger eigenvalue. This means all initial phases converge to the same one after enough long time. Because the larger eigenvalue is greater than 1, $\chi$'s total energy also increases gradually.

For $b^2<1$ i.e. $\cos^2\delta<1/\cosh^2\eta_m$, the two eigenvalues have unit magnitude and opposite arguments. Consequently there is no such phase convergence, and no overall change in $\chi$'s total energy should be expected.

Such difference is also quite clear from the figures. The boost rate has dependence on the initial phase and transforms a circle in the phase diagram to an ellipse. Because of such a transformation, the phases converge towards $x$-axis. On the other hand, $\delta$ rotates the ellipse and this may ruin the convergence if it's large enough. These two effects compete and, in consequence, the above two distinct behaviors of the system may occur.

This also explains the exact solution of \refeq{eom-chi}, the so-called Mathieu equation. The solution of Mathieu equation has bands of stable solutions in which the amplitude of $\chi$ generally remains at its initial scale as time passes, and bands of unstable solutions which have an exponentially growing $|\chi|$ w.r.t time. Those bands correspond exactly to the non-convergent and convergent cases respectively. 

For better understanding, we can also calculate the phase that is converged to. The fixed points under transformation matrix \refeq{pr-transmatrix} satisfy $y''/x''=y/x$, i.e.
\eq{\frac{y}{x}=\frac{y''}{x''}=\frac{xe^{2\eta_m}\sin\delta+y\cos\delta}{xe^{2\eta_m}\cos\delta-y\sin\delta}.}
The solutions are
\eq{\frac{y}{x}=\frac{\sinh\eta_m\pm\sqrt{\sinh^2\eta_m-\tan^2\delta}}{e^{\eta_m}\tan\delta}.\label{eq-pr-convergesol}}
\fig{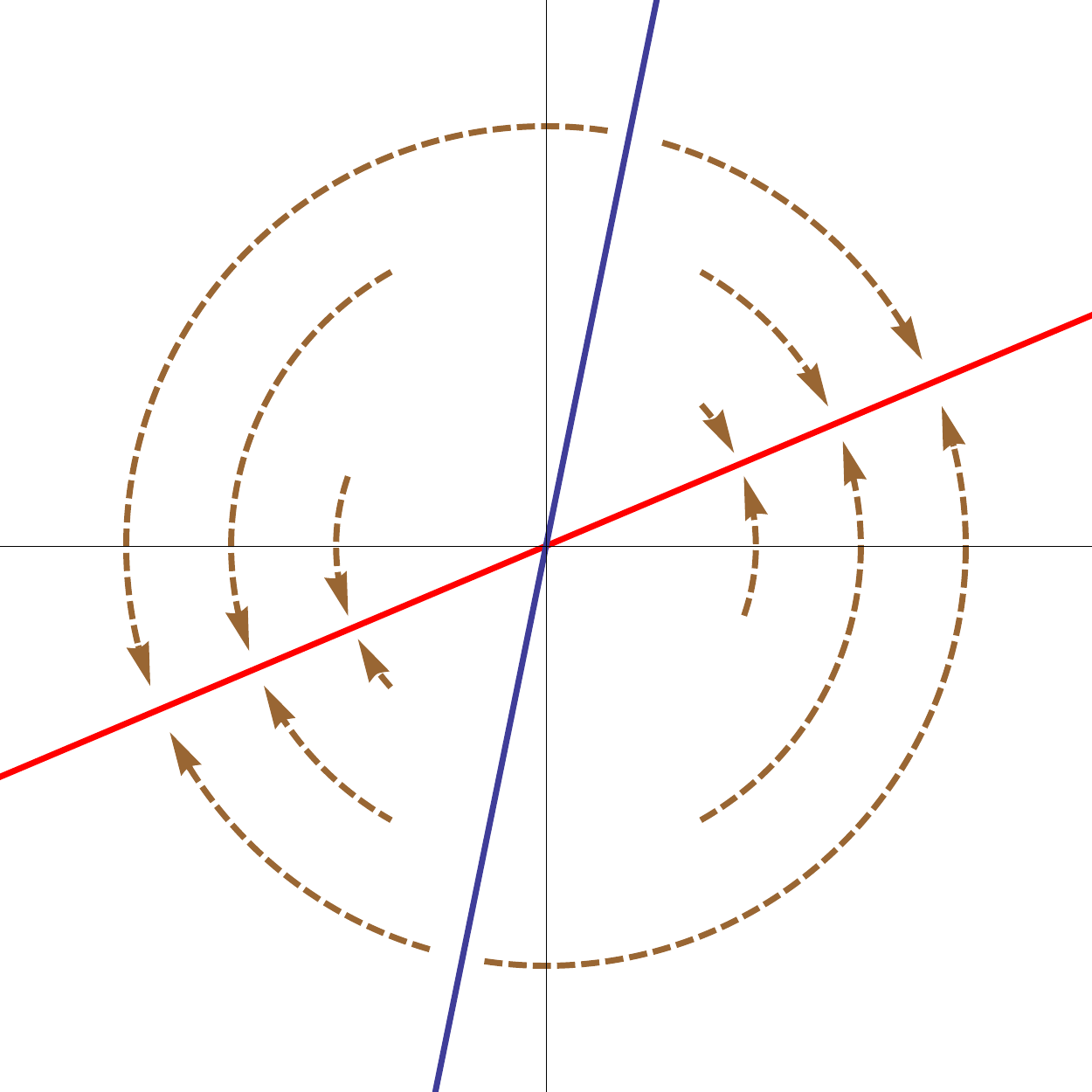}{The diagram showing the four fixed phases under transformation \refeq{pr-transmatrix}, and four regions of other phases with different directions converge towards. Here $\delta\approx\pi/6$. The red line is the phase converged to and is thus fixed under the transformation. The dashed curves with arrows show the directions other phases converge towards in different regions. The blue line stands at the boundaries of the neighboring regions heading towards opposite directions, and therefore is also invariant under the transformation. The red and blue lines are eigenstates with the larger and smaller eigenvalues respectively.}{\defaultfigsize}{t}

We can see how the solution is represented in the phase diagram in \refig{Graph-PhaseEvolveDirection.pdf}. From the figure we know at once we should take the minus sign in \refeq{pr-convergesol} as the phase converged to. Based on these results, we can further derive $\chi$'s long-time boost rate (which takes time far more than sufficient for convergence to take place, if there is). For the convergent case, the long time boost rate is simply the $\log$ of the larger eigenvalue in \refeq{pr-transmatrix}, and for the non-convergent case, no boost occurs so the long time boost rate is zero. We thus define the long time boost rate as
\eq{\eta_l(\delta)\equiv\left\{\begin{array}{l@{,\hspace{0.4in}}l}0&\displaystyle\cos^2\delta<\frac{1}{\cosh^2\eta_m}=\hf\\\displaystyle\log\Bigl(\cosh\eta_m\cos\delta+\sqrt{\cosh^2\eta_m\cos^2\delta-1}\,\Bigr)&otherwise.\end{array}\right.\label{eq-pr-etal}}
It's also visualized in \refig{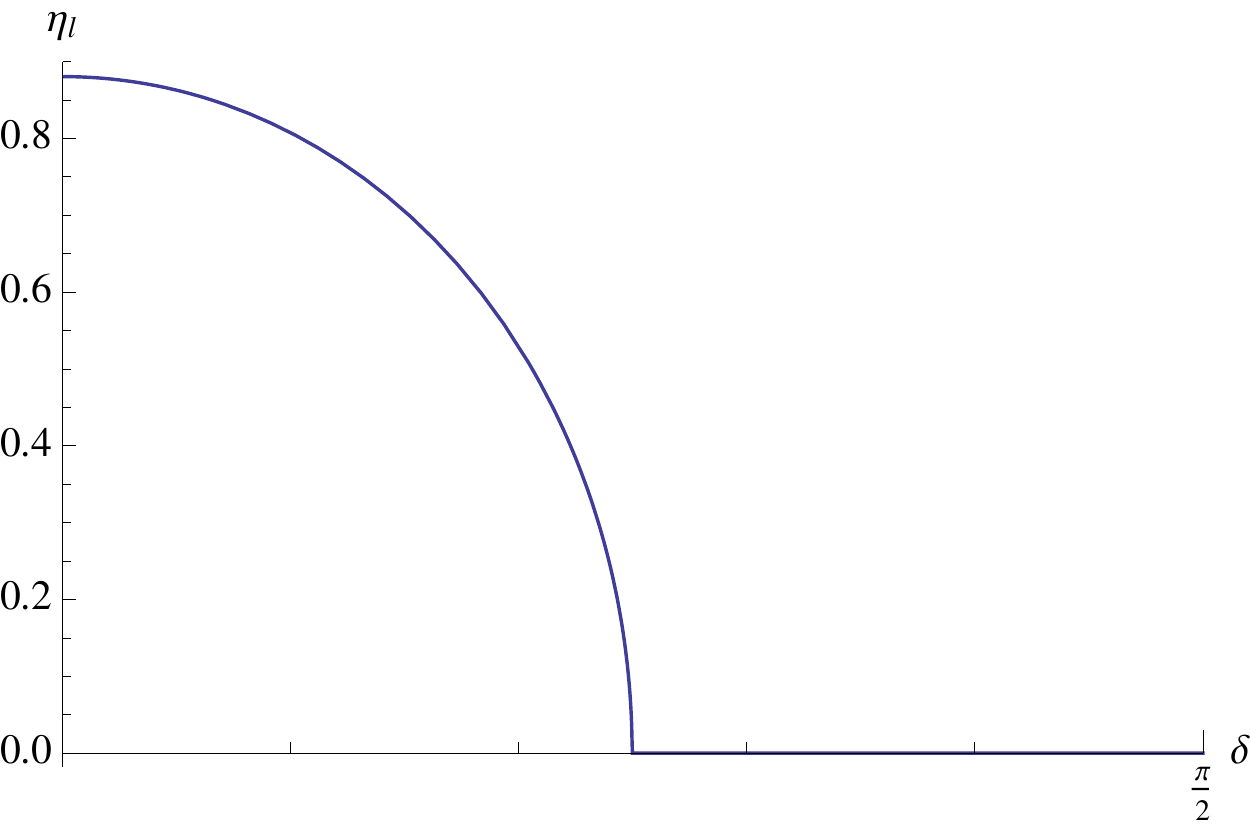}.
\fig{Graph-PhaseBoostRelationLong.pdf}{The long time boost rate under different rotation angles $\delta$. Smaller $\delta$ typically leads to larger boost rate because the eigenstate is more aligned with the phase of maximum boost rate. Large enough $\delta$'s don't generate any long-time boost effect because they are the non-convergent cases.}{\defaultfigsize}{t}

In \refig{Graph-PhaseBoostRelationLong.pdf}, we can see the convergent-non-convergent boundary is at $\pi/4$, which is exact and can be confirmed from the expression of $\eta_m$. This indicates only half of the $\delta$ values give a long-time boost effect. If we consider the case with a $\delta$ slowly scanning over $[0,2\pi)$ at a constant speed, we would get the average long-time boost rate
\eq{\langle\eta_l\rangle\equiv\frac{1}{2\pi}\int_0^{2\pi}\eta_l(\delta)\dd\delta\approx0.346.\label{eq-pr-etala}}
Please note there is relation $\eta_l(-\delta)=\eta_l(\pi+\delta)=\eta_l(\delta)$.

It's quite interesting here to notice $\langle\eta_l\rangle$ ``happens" to be very close to $\langle\eta\rangle$. Although the expressions of $\langle\eta\rangle$ and $\langle\eta_l\rangle$ are very different, this however is not a coincidence. $\langle\eta\rangle$ and $\langle\eta_l\rangle$ are actually identical and numerical integrals have confirmed that. It is not yet very clear why they are equal, but the following  might be a reasonable explanation. Every value of $\delta\in[0,2\pi)$ is actually hit the same number of times during the calculation of $\langle\eta\rangle$ and $\langle\eta_l\rangle$. Consequently the transformation matrices of all $\delta$'s operate on the initial state the same number of times, although the order of the operations is different for $\langle\eta\rangle$ and $\langle\eta_l\rangle$. The transformation matrices do not commute, but their commutation is only a rotation. The commutation could result in an either slightly larger or smaller boost rate, depending on whether the rotation is towards the major axis of the total transformation matrix or away from it. A huge number of such commutations is needed to make the two sequences the same. Some of the commutations contributes positively to the total boost rate while others negatively. When summed up, their effects cancel, and maybe one might be willing to think this as the reason of the equivalence between $\langle\eta\rangle$ and $\langle\eta_l\rangle$.

\secs{Steps Towards Cosmological Applications}
So far, we have established the framework to consider the boost effect of parametric resonance in a static background. To apply it practically on cosmology, there are still a few steps should be taken first. One is the need to put our framework into the expanding background; the other is the consideration of backreaction. We will discuss them in this section.

\ssecs{Dynamic Background}
When we apply this model in cosmology, a dynamic background is required.\footnote{Besides expansion, this model is also applicable in contracting phases without any prejudice. Because we are more concerned with the expanding phase, we will only discuss the effect of universe expansion here. The effect of contraction however, can be derived in the same way, and stages are simply reversely sequenced in most cases.} We will still neglect the backreaction in this section, so the effect of expansion on $\phi$ is simply a damping factor $|\phi|\propto a^{-\frac{3}{2}}(t)$, where $a(t)$ is the scale factor. The equation of motion of $\chi$ however becomes
\eq{\ddot\chi+3H\dot\chi+2\lambda\phi^2\chi=0,\label{eq-eb-eom-Hchi}}
where $H\equiv \dd a/a\dd t$ is the Hubble rate of the universe, and can be regarded as a constant for each half cycle of $\phi$. Writing $\chi=a^\frac{3}{2}\Psi$ simplifies \refeq{eb-eom-Hchi} to
\eq{\ddot\Psi+2\lambda\phi^2\Psi=0,\label{eq-eb-eom-Psi}}
in which we have neglected the effective mass contribution from universe expansion for $\lambda|\phi|^2\gg m^2>H^2$. Therefore we can see the universe expansion also gives $\chi$ an $a^{-\frac{3}{2}}$ term.

However the universe expansion has other impacts on our framework. As $|\phi|$ is damped by expansion, the rotation angle $\delta$ no longer remains constant across cycles of $\phi$. Therefore changes should be made to the calculations in \refssec{Long Time Evolution}. For convenience, we use the notation $x\subs{(n)}$ to indicate the variable $x$ of the $n$th half cycle of $\phi$, or at the time when $\phi$ reaches its maximum the $n$th time. When all the subscripts $\subs{(n)}$ are equal in one equation, we may neglect some or all of them, as long as no confusion might be caused. Primes here are defined as the difference
\eq{x'\subs{(n)}\equiv x\subs{(n+1)}-x\subs{(n)}.}

Different from the static version \refeq{pr-delta}, in an expanding background, $\delta\subs{(n)}$ should be redefined as
\eq{\delta\subs{(n)}\equiv\int_{t\subs{(n)}}^{t\subs{(n+1)}=t\subs{(n)}+\frac{\pi}{2m}}2\sqrt{2\lambda}|\phi|(t)\cos m t\dd t.}
To finish the integral, we manually fix $|\phi|(t)$ at $|\phi|\subs{(n)}$, so it gives the same result with \refeq{pr-delta}. This approximation however gives $\delta\subs{(n)}$ an error 
\eq{\Delta\delta\subs{(n)}<\int_{t\subs{(n)}}^{t\subs{(n+1)}}2\sqrt{2\lambda}||\phi|\subs{(n+1)}-|\phi|\subs{(n)}|\cos mt\dd t=\frac{2\sqrt{2\lambda}||\phi|'{}\subs{(n)}|}{m}}
where the outer $|\ |$ is to take absolute values. For short, we have
\eq{\Delta\delta<\frac{2\sqrt{2\lambda}||\phi|'|}{m}.\label{eq-eb-Deltadelta}}
Similarly we get
\eqa{\delta'&\equiv&\delta\subs{(n+1)}-\delta\subs{(n)}=\frac{2\sqrt{2\lambda}|\phi|'}{m},\label{eq-eb-deltap}\\
\delta''&\equiv&\delta'\subs{(n+1)}-\delta'\subs{(n)}=\frac{2\sqrt{2\lambda}|\phi|''}{m},\\
\Delta\delta'&<&\frac{2\sqrt{2\lambda}||\phi|''|}{m}=|\delta''|.\label{eq-eb-Deltadelta1}}

The universe expansion redshifts $|\phi|$, giving
\eq{\frac{|\phi|'}{|\phi|}=e^{-\frac{3\pi H}{2m}}-1=-\frac{3\pi H}{2m},}
which consequently redshifts $\delta$. The average effect of parametric resonance is then determined by the scales of $\delta$, $\delta'$ and $\delta''$.

When $\delta'\ll 1$, $\delta$ changes very slowly compared with the oscillation of $\phi$. The system therefore has enough time to converge to the eigenstate whenever there's any. Then the system can be regarded as trapped in its eigenstate which is slowly varying. The boost rate $\eta_l(\delta)$ changes correspondingly as $\delta$ slowly decreases, and the average boost rate in this case should then be taken as the long-time boost rate, $\langle\eta_l\rangle$, defined in \refeq{pr-etala}. As $\delta$ is slowly redshifted by universe expansion, the convergent and non-convergent cases are met in turn. This makes $\chi$'s evolution stair-like at this stage. Half of the time, the  amplitude of $\chi$ enjoys flat platforms lasting for cycles of $\phi$ during which it doesn't grow at all. The platforms are separated by steep growing regions which fills the other half of the time. The two distinct stages come in turn to get the stair-like curve of $\chi$'s amplitude.

When $\delta'\sim>1$ but $\delta''\ll 1$, $\delta$ varies fast but $\delta'$ varies very slowly. In this case, $\delta$ changes significantly between neighboring half cycles of $\phi$, but the amount changed stays almost fixed. For every half cycle of $\phi$, an almost constant $\delta'\sim1$ adds to the initial rotation angle $\delta$. Therefore for time long enough, the average effect is $\beta$ has an even probability at each value in $[0,2\pi)$. This corresponds to an even distribution of $\beta$ in $\eta(\beta)$ and, results in the average boost rate which should be chosen as $\langle\eta\rangle$ in \refeq{pdn-etaa}.

When $\delta''\sim>1$, the relevance between neighboring $\delta$'s or $\delta'$'s is small, so one usually considers $\delta\subs{(n)}$ as random in $[0,2\pi)$, being irrelevant with the history of evolution. When long enough time is experienced, the random distribution of $\delta$ gives the same results as the case of $\delta'\sim>1$ and $\delta''\ll 1$ so $\langle\eta\rangle$ should also be chosen as the average boost rate.

In realistic cases however, it is rare to tell the last two stages apart. They share the same average boost rate $\langle\eta\rangle$ and most of their characteristics are alike, so there is no need to distinguish them. Moreover, the second category has a $\delta''\ll1$ but non-vanishing. To demonstrate the difference between the last two categories, the even probability of $\beta$ needs to be shown. This however requires a prolonged constant value of $\delta'$, which in most cases is very difficult to achieve with a non-vanishing $\delta''$, even though it's much smaller than 1. For these reasons, we will treat them as a whole ($\delta'\sim>1$) in our following discussions and call it the \emph{large $\delta'$ stage}, in which $\delta\subs{(n)}$ is randomly distributed and independent of the history. In the same way, we name $\delta'\ll 1$ as the \emph{small $\delta'$ stage}.

We now fall back to check the approximation we made previously in \refeq{eb-Deltadelta} and \refeq{eb-Deltadelta1}, where $|\phi|$ and $|\phi'|$ are regarded as constants in every half cycle. They give errors $\Delta\delta<|\delta'|$ and $\Delta\delta'<|\delta''|$, which fit perfectly with the three stages above. Therefore this approximation is applicable here.

\fig{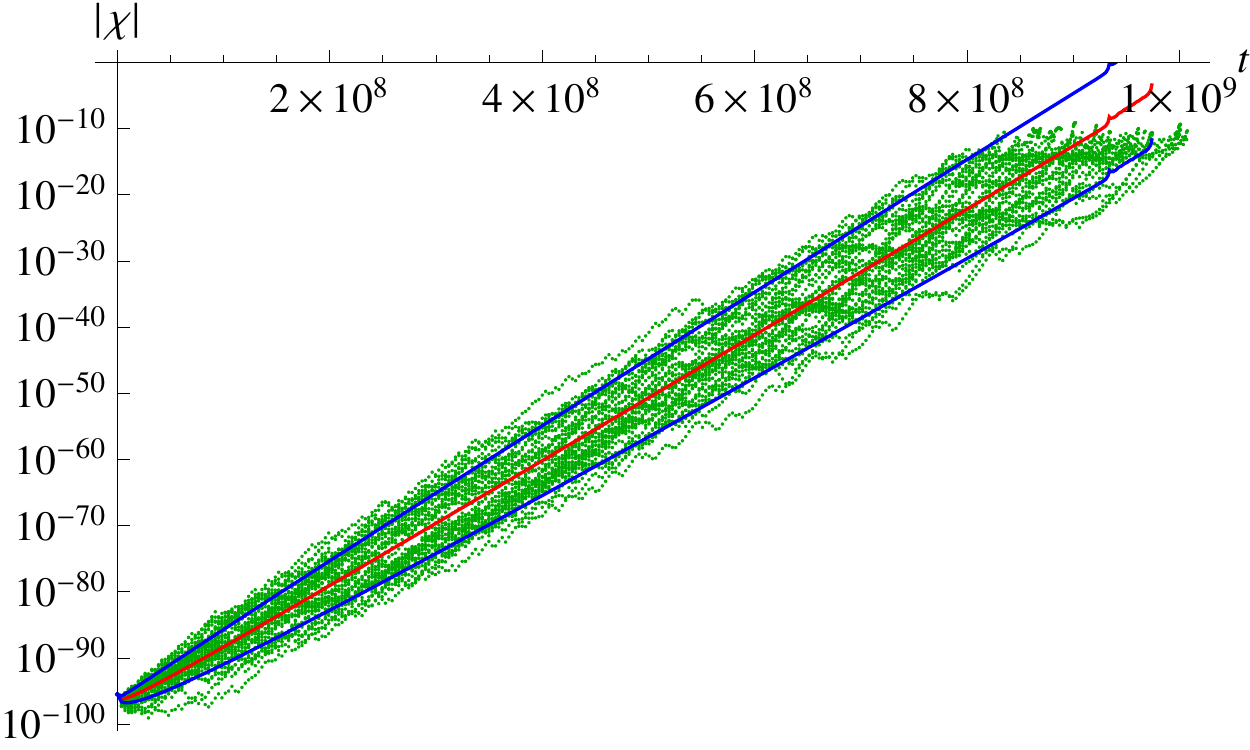}{The numerical simulation result of many preheating processes compared with our statistical estimation. (See \refssec{Preheating}.) Universal parameters are $m=10^{-6}\mpl$, $M=10^{-10}\mpl$, and $\gamma=10^{-9}$. The only varying parameter for different simulations is the interaction strength $\lambda$, starting at $10^{-4}$ and ending at $4\times10^{-3}$, with a step increment $10\%$. The green plots are the data from numerical simulation, taken every time $\phi$ reaches maximum. The red line is the statistical average result and the two blue lines enclose the area within error $1\sigma$, derived from $\Delta\langle\acute\eta\subs{(n)}\rangle$ in \refeq{db-daeta}. On the left half, we see a reasonable amount of plots lying outside the blue curves because we only expect $68\%$ of the plots in $1\sigma$. On the right half the universe actually enters the small $\delta'$ stage and the error doesn't grow so much. However the blue lines are derived under the assumption that the whole evolution is the large $\delta'$ stage, so they are only functions of $n$ and keep extending. Thus we find on the right half, the number of plots outside the blue lines is significantly decreased. The numerical data stop when backreactions start to be important.}{\defaultfigsize}{t}
Before we end this discussion on dynamic background, we would like to briefly calculate the statistical results from the random $\delta$ for the large $\delta'$ stage. In $\frac{n}{2}$ cycles of $\phi$, the boost effect of parametric resonance occurs $n$ times. The total boost rate is
\eq{\acute\eta\subs{(n)}(\delta)\equiv\sum_{i=1}^n\eta(\delta\subs{(i)}).}
Assuming absolute randomness of all $\delta\subs{(i)}\in[0,2\pi)$, we get the expectation value of $\acute\eta\subs{(n)}(\delta)$ as
\eq{\langle\acute\eta\subs{(n)}\rangle\equiv\frac{1}{(2\pi)^n}\int\dd^n\delta\sum_{i=1}^n\eta(\delta\subs{(i)})=\frac{n}{2\pi}\int_0^{2\pi}\eta(\delta)\dd\delta=n\langle\eta\rangle\approx0.346\,n.\label{eq-db-aeta}}
This is obvious because it's just how $\langle\eta\rangle$ is defined. To further compute the statistical error, we calculate
\eqa{\langle\acute\eta^2\subs{(n)}\rangle&\equiv&\frac{1}{(2\pi)^n}\int\dd^n\delta\sum_{i=1}^n\eta^2(\delta\subs{(i)})\nonumber\\
&=&\frac{n}{2\pi}\int_0^{2\pi}\eta^2(\delta)\dd\delta+\frac{n(n-1)}{4\pi^2}\Bigl(\int_0^{2\pi}\eta(\delta)\dd\delta\Bigr)^2\nonumber\\
&=&n\langle\eta^2\rangle+n(n-1)\langle\eta\rangle^2.}
and thus the statistical error can be estimated as
\eq{\Delta\langle\acute\eta\subs{(n)}\rangle\equiv\sqrt{\langle\acute\eta^2\subs{(n)}\rangle-\langle\acute\eta\subs{(n)}\rangle^2}=\sqrt{n}\sqrt{\langle\eta^2\rangle-\langle\eta\rangle^2}\approx0.539\sqrt n.\label{eq-db-daeta}}
To justify both the statistical calculation and the definition of the large and small $\delta'$ stages, here we also give \refig{Graph-ReheatingError.pdf} through multiple numerical simulations.

\ssecs{Backreaction}
In this section, we discuss the backreaction of $\chi$ by taking into account the $\lambda\chi^2$ term in $\mphi^2$, the effective mass squared of $\phi$. Since we assumed homogeneity, we will not discuss other backreactions here. In most cases of parametric resonance, the assumption $\lambda\chi^2\ll m^2$ starts valid. However as $\chi$'s amplitude gets boosted by resonance, this assumption would finally break at $|\chi|\sim m/\sqrt\lambda$, provided the boost effect hasn't yet been halted by other causes. Because $|\chi|^2\propto\mchi^{-1}\approx1/\sqrt{2\lambda\phi^2}$, for every half cycle of $\phi$, $|\chi|$ reaches its maximum when $\phi\rightarrow0$. From this point of view, the violation of this approximation first starts small --- only at $\phi\rightarrow0$ --- and then expands to larger $\phi$ gradually while $|\chi|$ is being further boosted.

From the relation $|\chi|^2\propto1/\sqrt{2\lambda\phi^2}$, we can find the effective potential for $\phi$, which is actually
\eq{V\subs{eff}(\phi)\equiv\hf m^2\phi^2+\lambda\phi^2\chi^2=\hf m^2\phi^2+\hf\lambda|\chi|_m^2|\phi||\phi|\subs{abs},} 
where $|\chi|_m$ is $\chi$'s amplitude at maximum $\phi$, and $|\ |\subs{abs}$ means taking the absolute value (and $|\ |$ still indicates taking the amplitude). Here we can see the additional part of $V\subs{eff}(\phi)$ is proportional to the absolute of $\phi$. Compared with $\hf m^2\phi^2$ which is proportional to $\phi^2$, the effective potential from interaction $V\subs{eff}(\phi)$ agrees with our previous instinct that the backreaction first comes to important at small $\phi$. 

So we consider the motion of $\phi$ for half a cycle, starting from $t=0$, $\phi=-|\phi|$, $\dot\phi=0$, and $|\chi|=|\chi|_m$. We will use $\gamma^2\equiv\lambda|\chi|_m^2/m^2$ for simplicity. When $\phi<0$, $\phi$'s equation of motion is
\eq{\ddot\phi+m^2\phi-\lambda|\chi|_m^2|\phi|=0.}
We can analytically solve this equation and arrive at $\phi=0$. The solution is
\eq{t=t_1\equiv\frac{1}{m}\arccos\frac{\gamma^2}{\gamma^2+1},}
\eq{\dot\phi(t=t_1)=(\gamma^2+1)m|\phi|\sin mt_1.}

At $\phi=0$, $\chi$ is boosted by parametric resonance. This gives the interaction part of $V\subs{eff}(\phi)$ a factor of $e^{2\eta}$. Here $\eta$ is still the boost rate. For this reason, the equation of motion is slightly modified for $\phi>0$,
\eq{\ddot\phi+m^2\phi+\lambda|\chi|_m^2|\phi|e^{2\eta}=0.}
We then solve it again to arrive at $\dot\phi=0$ and finish the half cycle. At the $\phi>0$ side when $\phi$ reaches maximum again at time $t=t_2$, we have
\eq{t_2=t_1+\frac{\pi}{2m}-\frac{1}{m}\arctan\frac{\gamma^2e^{2\eta}}{\sqrt{\gamma^2+1}},}
\eq{\phi(t=t_2)=|\phi|(\sqrt{\gamma^4e^{4\eta}+2\gamma^2+1}-\gamma^2e^{2\eta}),}
and
\eq{|\chi|=e^\eta|\chi|_m\frac{|\phi|}{\phi}.}
We can clearly see when $\eta=0$, there is no boost effect and the motion of the system becomes symmetric w.r.t $\phi=0$.

When the backreaction is weak, i.e. $\gamma^2\ll 1$, we can expand the results to first order and get
\eqa{t_2&=&\frac{\pi}{m}-\frac{e^{2\eta}+1}{m}\gamma^2,\\
\phi(t_2)&=&|\phi|\Bigl(1-(e^{2\eta}-1)\gamma^2\Bigr),\\
|\chi|&=&e^\eta|\chi|_m\Bigl(1+(e^{2\eta}-1)\gamma^2\Bigr),\\
\delta&=&\frac{2\sqrt{2\lambda}|\phi|}{m}\biggl(1-\Bigl(\frac{\pi}{2}(e^{2\eta}+1)-2\Bigr)\gamma^2\biggr).\label{eq-p0c-delta}}
Therefore, in case the backreaction presents, the oscillation of $\phi$ becomes faster and that of $\chi$ is compared slower. The non-vanishing $\eta$ gives a boost effect. Amplitudes $\phi(t_2)$ and $|\chi|(t_2)$ are thus affected and become slightly different with their original values at $t=0$, suggesting an energy transfer.

Whenever one takes this backreaction into account, one should adopt these corrections. However for weak backreactions in which $\gamma^2\ll1$, the corrections actually can be neglected. Such backreaction only comes to important when $\gamma^2\sim>0.1$. In some cases, however, the exponential boost effect is terminated by other causes before $\gamma^2$ reaches $0.1$, so it never comes to important.

If the system does evolve to $\gamma^2\sim>0.1$, one then needs to take it seriously. The most significant effect is $\phi$ oscillates faster and faster, due to the effective mass contribution from the backreaction. This has several consequences. First and most obviously, because the boost rate is counted per half cycle of $\phi$, a faster rate of $\phi$'s oscillation also leads to a faster boost to $\chi$. The faster boost acts back on $\phi$'s effective mass and causes $\phi$ to oscillate even faster. The positive feedback gives an explosive boost effect to $\chi$ once the backreaction comes to important. The explosive boost effect is soon shut down when $|\chi|_m$ reaches the same scale with $|\phi|$. This is because $\phi$ and $\chi$ would then have similar frequencies and the approximate replacement of $\chi^2$ with $\langle\chi^2\rangle$ would be no longer applicable. This immediately ends the exponential boost effect of parametric resonance.

Moreover, this backreaction may also offer a major contribution to $\delta'$. Here we give a simple estimate based on \refeq{p0c-delta}. By definition, we have $(\gamma^2)'=2\lambda(|\chi|_m^2)'/m^2=\gamma^2(e^{2\eta}-1)$. So the correction to $\delta'$ up to first order of $\gamma^2$ is
\eq{\Delta\delta'=-\frac{2\sqrt{2\lambda}|\phi|\gamma^2}{m}\Bigl(\frac{\pi}{2}(e^{4\eta+2\eta'}-1)-e^{2\eta}+1\Bigr).\label{eq-br-deltap}}
When it has the same magnitude with the originally defined $\delta'$ in \refeq{eb-deltap} in an expanding universe, we can solve the magnitude of $\gamma^2$,
\eq{\gamma^2\sim\frac{H}{m}<0.1.}
This means the impact of backreaction on rotation angle $\delta$ is even more noticeable than that on $\phi$'s period.

In the same way, it's easy to derive $\Delta\delta'$ has unit magnitude when $\gamma^2\sim m/\sqrt\lambda|\phi|\ll1$. Normally in an expanding universe where the backreaction is negligible, $|\delta'|$ is always decreasing so the resonance would gradually undergo a shift from the large $\delta'$ stage to the small $\delta'$ stage. The relation of $\gamma^2$ however tells that the parametric resonance may be subject to a shift back to the large $\delta'$ stage from the small $\delta'$ stage, or even, the small $\delta'$ stage may not appear at all. The back shift, whenever exists, always lies ahead of the explosive boost. So for deep discussions of the ending conditions of the exponential boost, one should always take it into account.

\secs{Cosmological Applications}
\ssecs{Preheating}
Preheating is a possible stage after inflation, during which the inflaton decays into particles through parametric resonance. Preheating was first realized in \cite{Kofman:1994rk,Kofman:1997yn}, and the parameter space is the same with what we have discussed. Therefore, during preheating, the number of preheated particles grows exponentially, much faster than the redshift rate from universe expansion.

Although preheating should be followed by particle thermalization, and possibly a second stage of reheating between them, they are not within our consideration in this paper. Instead of calculating the reheating temperature and how long reheating and thermalization last, this section is devoted to showing how this framework may be applied on specific preheating calculations. Details about the second stage of reheating and thermalization can be found in \cite{Kofman:1997yn} and \cite{Davidson:2000er}. For the same reason, we don't care the reheating ratio between the dark and visible sector. To solve the dark radiation problem, one might want to refer to inflation models with more particle physics foundation, e.g.\ SUSY inspired inflation and the subsequent reheating and thermalization studies\cite{Allahverdi:2011aj, Allahverdi:2007zz, Mazumdar:2010sa}.

During preheating, the universe is dominated by $\phi$ in our model, which is usually the inflaton. As a demonstration, we here just consider the simple case of the preheating of a generic slowroll inflation with inflaton $\phi$. To solve problems like horizon and flatness, inflation requires an e-folding at least $N\approx60$. At the beginning of such slowroll inflations, there should be $\lambda\chi^2\ll m^2$. Otherwise there would be a significant decrease in $\phi$'s effective mass and thus no inflation although $\phi$ would still be slowrolling.

In general, the end of slowroll inflation is signaled by the destruction of the slowroll condition of $\phi$, after which $\phi$ begins to oscillate and preheating takes place. We would like to neglect the first cycle or so of $\phi$ in the beginning of preheating here, so $\mphi\gg H$ holds for the rest of the preheating period. Because $\lambda\chi^2\ll m^2$ holds in the beginning of inflation, and $\chi$ is redshifted greatly during inflation, this relation is expected to still hold for a very long period in preheating. Because the backreaction terminates parametric resonance once it comes to important, it acts as an ending condition of the preheating scenario. Therefore we can first neglect the presence of backreaction, and then calculate the take-over of backreaction separately. 

We label the beginning of inflation as $0$, the end of inflation (i.e. the start of preheating) as $1$, and the point we start to consider in preheating as $2$. The evolution of the universe is then divided by such points, whose labels we use as subscripts, into intervals such as $1\rightarrow2$ for the convenience of future references. To be exact, we define the point $2$ as
\eq{|\phi_2|\equiv e^{-1}|\phi_1|.}
$\chi$ is redshifted during inflation ($0\rightarrow1$), but the boost and redshift effect approximately cancel during $1\rightarrow2$. We therefore get
\eq{|\chi_2|\sim|\chi_1|=e^{-\frac{3}{2}N}|\chi_0|.}
The universe is dominated by $\phi$ during preheating so the Hubble rate writes
\eq{H^2=\frac{4\pi m^2|\phi|^2}{3\mpl^2}}
in which for point 1 and after,
\eq{|\phi|=|\phi_1|\Bigl(\frac{a_1}{a}\Bigr)^\frac{3}{2}.\label{eq-rh-phi}}

After point 2, the system starts at the large $\delta'$ stage. There may be a transition from large to small $\delta'$ stage during the evolution, but the termination of parametric resonance by backreaction is certain to take place at the large $\delta'$ stage, which is ensured by its influence on $\delta'$ shown in \refeq{br-deltap}, slightly before the positive feedback ends the exponential boost. We thus also label the point after which the large $\delta'$ is ensured by backreaction, and the point backreaction becomes important and consequently terminates parametric resonance as points 3 and 4, also in chronological order.

\cm{

Substituting them into \refeq{pr-delta} and \refeq{eb-deltap} gives
\eqa{\delta&=&\frac{2\sqrt\lambda|\phi|}{m},\\
\delta'&=&-\frac{\sqrt{3\pi\lambda}\,2\pi|\phi|^2}{m\mpl}.}

We use $\delta'<e^{-2}$ as the precise criterion of $\delta'\ll1$, so the small $\delta'$ stage corresponds to $|\phi|<\Bigl(\frac{m\mpl}{\sqrt{3\pi\lambda}\,2\pi}\Bigr)^\frac{1}{2}e^{-1}$. Similarly, for the large $\delta'$ stage, we have $|\phi|>\Bigl(\frac{m\mpl}{\sqrt{3\pi\lambda}\,2\pi}\Bigr)^\frac{1}{2}e^{-1}$. We thus label the point of this transition as 3, i.e. define point 3 as
\eq{|\phi_3|\equiv\Bigl(\frac{m\mpl}{\sqrt{3\pi\lambda}\,2\pi}\Bigr)^\frac{1}{2}e^{-1}.}
Then the period $2\rightarrow3$ is the large $\delta'$ stage, where the boost effect has a random signature. And for the time after point $3$, small $\delta'$ applies and the stair-like boost rates can be found.
}

After defining the timeline, we now work on the evolution of $\chi$. The large and small $\delta'$ stages share the same average boost rate $\langle\eta\rangle$, so after point 2, field $\chi$ is affected by parametric resonance and becomes proportional to $e^{\frac{\langle\eta\rangle}{\pi}m(t-t_2)}$. Due to universe expansion, $|\chi|$ is redshifted and thus acquires the factor $(a_2/a)^\frac{3}{2}$. As we have demonstrated in \refssec{Rolling Stage}, $|\chi|\propto\mphi^{-\hf}$, this gives $|\chi|$, from the redshift of $\phi$ field, another factor $(a/a_2)^\frac{3}{4}$. We multiply them all together to get the statistically average evolution of field $\chi$
\eq{|\chi|=|\chi_2|\Bigl(\frac{a_2}{a}\Bigr)^\frac{3}{4}e^{\frac{\langle\eta\rangle}{\pi}m(t-t_2)}.}

In a matter-like universe, we have for any time $t_a$ and $t_b$,
\eq{\biggl(\frac{a(t_b)}{a(t_a)}\biggr)^\frac{3}{2}=1+\sqrt\frac{6\pi\rho(t_a)}{\mpl^2}\,(t_b-t_a),}
which, under the condition $\log a(t_b)-\log a(t_a)>1$, can be simplified to
\eq{t_b-t_a=\biggl(\frac{a(t_b)}{a(t_a)}\biggr)^\frac{3}{2}\frac{\mpl}{\sqrt{6\pi\rho(t_a)}}\label{eq-p-timeint}}
where $\rho(t_a)$ is the total energy density of the universe at time $t_a$. Because $\phi$ dominates preheating, the evolution of $|\chi|$ finally writes
\eq{|\chi|=|\chi_2|\biggl(\frac{a_2}{a}\biggr)^\frac{3}{4}e^{\frac{3}{2}B(\frac{a}{a_2})^\frac{3}{2}},\label{eq-p-chievolv}}
in which
\eq{B\equiv\frac{2\langle\eta\rangle\mpl}{(3\pi)^\frac{3}{2}|\phi_2|}.}

From the above equation, we immediately see $|\chi|$ is actually independent of $\lambda$ and $m$. It's obvious $|\chi|$ is $\lambda$-independent from the constant boost rate. The dependence on $m$, the bare mass of $\phi$, is however canceled by the same $m$-dependence of $\rho$. This can be also viewed as $m$ defines a time scale which both the Hubble rate and the boost rate of $\chi$ per time interval are proportional to.

We then derive the evolution of $\chi$ at points 3 and 4 which are defined from backreaction. Here we assume $\phi$ is the inflaton of a generic slowroll inflation of $N$ e-folds. At the beginning of it, i.e.\ point 0, there should be $\lambda|\chi_0|^2<m^2$ to allow inflation to happen. After being redshifted all along inflation, $|\chi_1|$ then becomes extremely small and this allows a safe period of preheating before the backreaction becomes important. After that, backreaction may get strong and, points 3 and 4 may be met in turn at any moment. To show the relation $\lambda|\chi_0|^2<m^2$, we deploy the parameter $\gamma_0^2\equiv\lambda|\chi_0|^2/m^2<1$. Here we want to give a conservative and natural estimation, so we assume $-100<\log\gamma_0<0$.

As we have demonstrated in \refssec{Backreaction} that backreaction ensures the large $\delta'$ stage at $\gamma^2\sim> m/\sqrt\lambda|\phi|$, we therefore define point 3 as
\eq{\frac{\lambda|\chi_3|^2}{m^2}=\gamma_3^2\equiv\frac{m}{\sqrt\lambda|\phi_3|}.}
Given the evolution of both fields, \refeq{rh-phi} and \refeq{p-chievolv}, this can be reduced to the equation of $\nxx{2}{3}$, the e-folding the universe experienced between point 2 and 3,
\eq{Be^{\frac{3}{2}\nxx{2}{3}}-\nxx{2}{3}=N+\frac{1}{6}\log\frac{m^2}{\lambda|\phi_2|^2\gamma_0^4}.\label{eq-rh-n23}}

Also, we can work on point 4 in the same way. After defining point 4 as $\lambda|\chi_4|^2\equiv m^2$, the calculation is straightforward to give
\eq{Be^{\frac{3}{2}\nxx{2}{4}}-\hf\nxx{2}{4}=N-\frac{2}{3}\log\gamma_0.\label{eq-rh-n24}}

In both of the above equations, the dominant term on the l.h.s is the exponential one. We can thus see the number of e-folds of preheating is only very weakly dependent on $\lambda$ and $\gamma_0$, because they are inside two ``log''s. This is because $\phi$ oscillates faster and faster as the universe expands, and in consequence $\chi$ acquires a much larger boost rate per Hubble time $H^{-1}$.

\fig{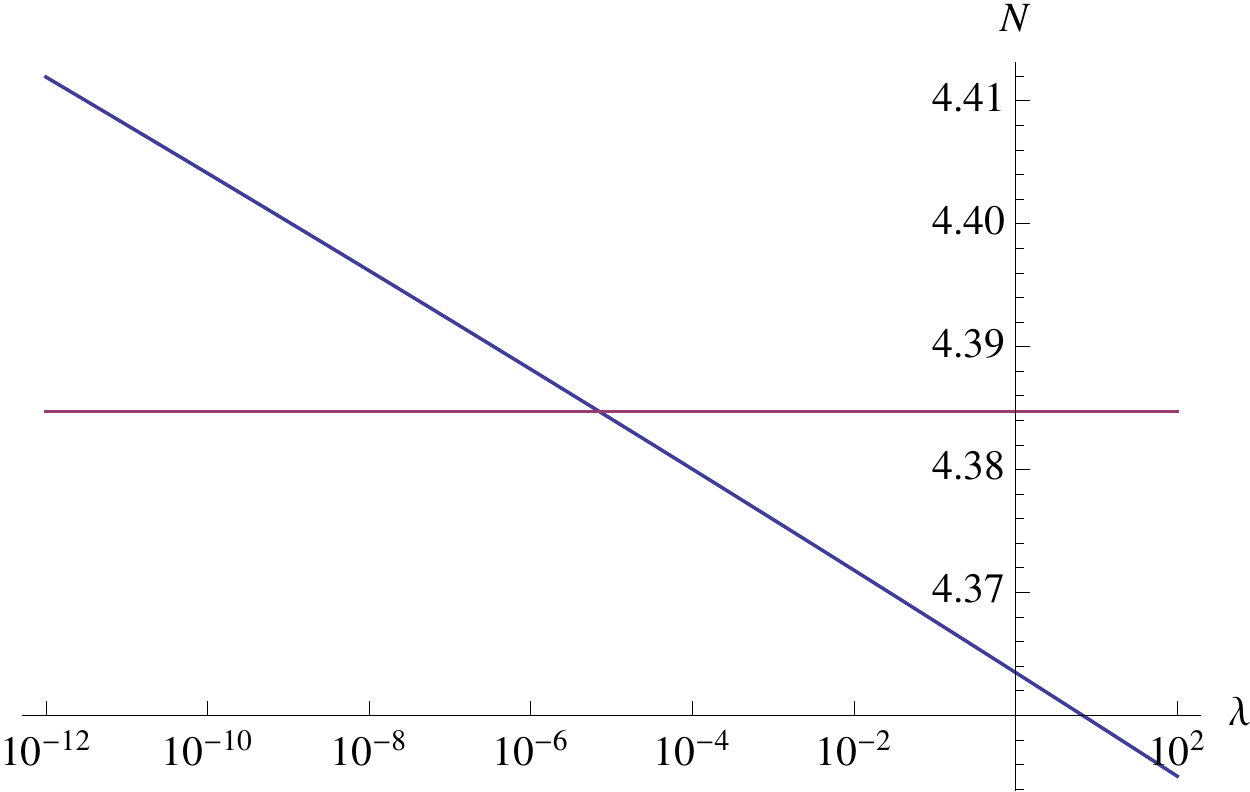}{$\nxx{2}{3}$ and $\nxx{2}{4}$ as functions of $\lambda$. Blue and purple correspond to $\nxx{2}{3}$ and $\nxx{2}{4}$ respectively. It's easy to see on the figure that $\nxx{2}{3}$ is only weakly dependent on $\lambda$ and $\nxx{2}{4}$ is independent of $\lambda$. Parameters are chosen to accord with the quadratic slowroll inflationary model.}{\defaultfigsize}{t}
To provide a brief example, we take the simple case of quadratic slowroll inflation which ends at $|\phi_1|^2=27\mpl^2/16\pi$, and has the number of e-folds of inflation $N=60$. We also need the value $m\sim 10^{-6}\mpl$, which comes from the strength of observed CMB anisotropy. \Refig{Graph-ReheatingNLargeR.pdf} shows the relations $\nxx{2}{3}$ and $\nxx{2}{4}$ have with $\lambda$ at constant $\gamma_0=0.1$. The dependence of $\gamma_0$ is weak, and a smaller $\gamma_0$ simply shifts both $\nxx{2}{3}$ and $\nxx{2}{4}$ upwards a bit with the same amount. For $\gamma_0=e^{-100}$, there is $\nxx{2}{4}=4.86$.

From \refeq{rh-n24}, we can also calculate the value of the r.h.s from $\nxx{2}{4}$. For $\nxx{2}{4}=4$ and 5 respectively, there is $N-\frac{2}{3}\log\gamma_0=32$ and 155. As a result, for most cases where $N$ is not very large and $\gamma_0$ is not too small, preheating is typically ended by backreaction at the 5th e-fold of parametric resonance i.e.\ the 6th e-fold after inflation, provided no other mechanism terminates preheating before that.

In \refssec{Backreaction}, we have suggested the correction from backreaction to $\delta'$ always comes to important earlier than that to $\mphi$. In \refig{Graph-ReheatingNLargeR.pdf}, it agrees with large $\lambda$. For small $\lambda$, the figure suggests a reverse in their sequence. The reverse however actually reflects the violation of $\lambda|\phi|^2\gg m^2$. For values of $\lambda$ on the left of the intersection point of the two curves in \refig{Graph-ReheatingNLargeR.pdf}, the condition $\lambda|\phi|^2\gg m^2$ should be broken well ahead of point 4 being reached, so the behavior of the system afterwards in fact can't be predicted by our theory. This indicates another possible exit of parametric resonance, which is caused by a small $|\phi|$ that is comparable to $m/\sqrt\lambda$. We label this exit as subscript $a$, and from the relation $\lambda|\phi_a|^2\equiv m^2$, we can easily derive the e-folding
\eq{\nxx{2}{a}=\frac{1}{3}\log\frac{\lambda|\phi_2|^2}{m^2}.}

Since both point $a$ and $4$ act as exits of parametric resonance, it is important to know which point comes earlier to find out by who and when parametric resonance is terminated. By setting $\nxx{2}{4}=\nxx{2}{a}$, calculations give a critical value for $\lambda$
\eq{\lambda_a\approx\frac{(N-\frac{2}{3}\log\gamma_0)^2m^2}{B^2|\phi_2|^2}\sim10^{-5}.}
For $\lambda>\lambda_a$, parametric resonance ends with backreaction. The amplitudes of $\phi$ and $\chi$ reach the same scale of magnitude in the end and equilibrium may be reached between the two fields. For $\lambda<\lambda_a$ however, parametric resonance is terminated by the violation of $\lambda|\phi|^2\gg m^2$. In this case, the evolution afterwards is not characterized by our theory, and we only know at the time this condition is broken, equilibrium has not yet been reached. Moreover, for $\lambda<m^2/|\phi_2|^2$, there is $\nxx{2}{a}<0$. This indicates the condition of our framework is not satisfied and such a parametric resonance does not take place immediately after inflation.

Meanwhile, we are also interested in the transition point from the large $\delta'$ stage to the small $\delta'$ stage. In the context of preheating, $\delta'$ is defined as
\eq{\delta'\equiv-\frac{2\pi^\frac{3}{2}\sqrt{6\lambda}|\phi|^2}{m\mpl}.}
When $\delta'$ reaches $e^{-2}\sim0.1$, we think the transition takes place and label it as point $b$. We thus derive the e-folding from point 2 to $b$
\eq{\nxx{2}{b}=\frac{1}{3}\log\frac{2e^2\pi^\frac{3}{2}\sqrt{6\lambda}|\phi_2|^2}{m\mpl}.}

For $\nxx{2}{b}>\nxx{2}{a}$, parametric resonance is terminated before the transition point, so the whole process is large $\delta'$. Similarly if $\nxx{2}{b}>\nxx{2}{3}$, the large $\delta'$ stage is also guaranteed so no small $\delta'$ stage takes place. For $\nxx{2}{b}<0$, parametric resonance starts from the small $\delta'$ stage, and whether a large $\delta'$ stage presents (although very short) depends on how preheating is terminated.

\fig{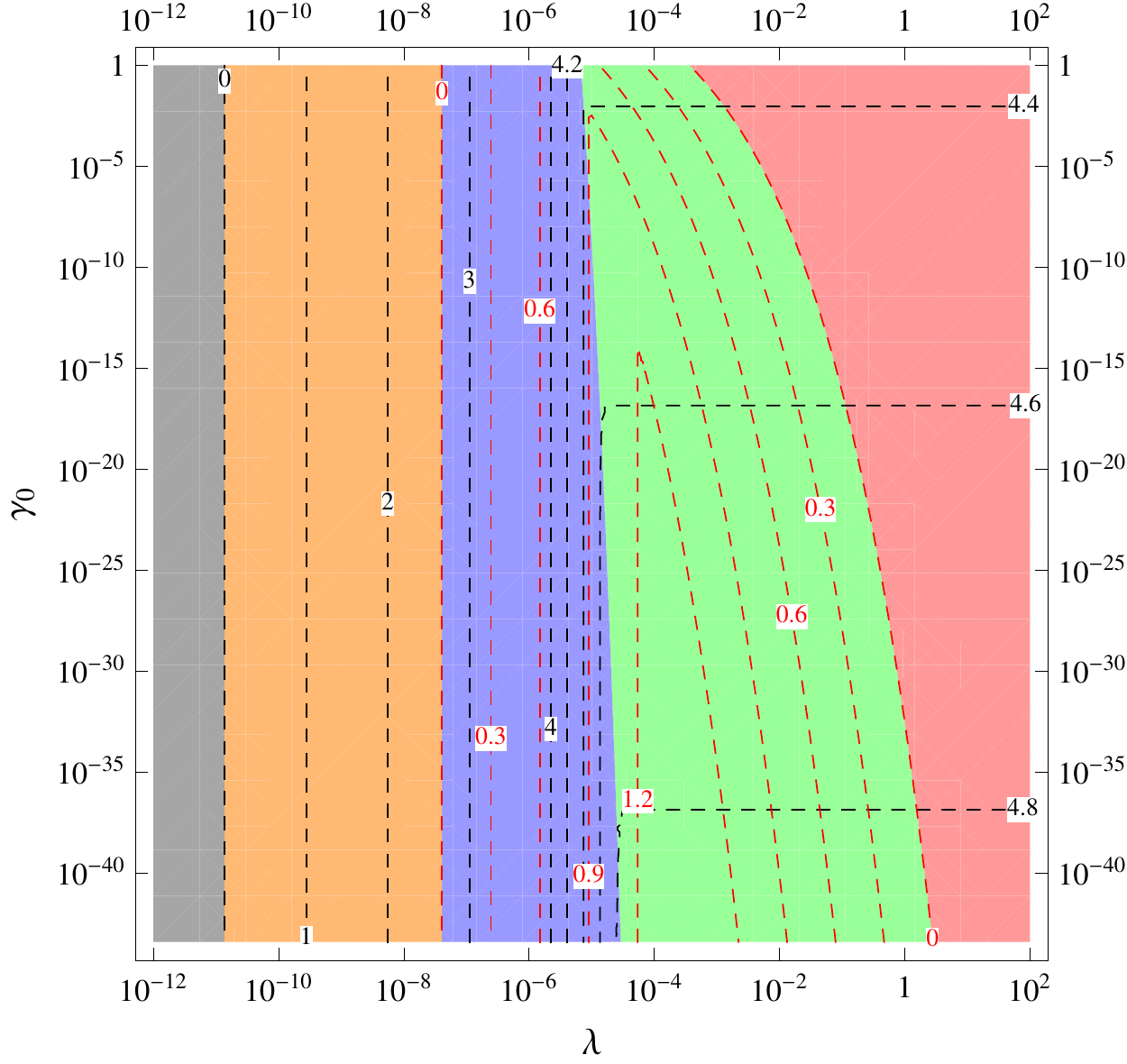}{The different circumstances of preheating in the parameter space of $\lambda$ and $\gamma_0$. The green and red regions are backreaction termination with and without the small $\delta'$ stage respectively. The blue and orange regions are terminated by the small $|\phi|$, with and without the small $\delta'$ stage correspondingly. In the gray region, the preheating stage is not entered after inflation. Among the dashed lines, the number of total e-folds of parametric resonance is contoured in black and that of the small $\delta'$ stage is contoured in red.}{0.8}{t}
To conclude this subsection, we perform a more precise calculation of the parametric resonance of quadratic slowroll inflation, and give \refig{Graph-ReheatingPSpace.pdf} which divides the parameter space of $\lambda$ and $\gamma_0$ into five sections. The small $\delta'$ stage may either exist or not exist for the terminations by both backreaction and the small $|\phi|$. There is also the case that no parametric resonance shall take place within our framework, so in all there are five sections. The numbers of the total e-folds of parametric resonance and the e-folds of the small $\delta'$ stage are also contoured in \refig{Graph-ReheatingPSpace.pdf}. For the termination by small $|\phi|$ however, the reheating is incomplete and there should be a subsequent decay of inflaton after the termination of parametric resonance, but our theory is unable to characterize it. From the figure, we also confirm that if preheating is terminated by backreaction, it should typically happen at the 5th e-fold of preheating, i.e.\ the 6th e-fold after inflation.

\ssecs{Locked Inflation}
Locked inflation, or new old inflation\cite{Dvali:2003vv}, is a fastroll inflation scenario during which a fastroll field $\phi$ locks the other field $\chi$ at its false vacuum. The false vacuum has a nonzero energy density which drives inflation. Locked inflation has the same potential shape with hybrid inflation\cite{Linde:1993cn}, but the parameter configuration is different.

Not long after the proposal of locked inflation, however, opposite opinions are posted suggesting underlying problems\cite{Copeland:2005ey,Easther:2004qs}. In \cite{Copeland:2005ey}, they concluded that saddle inflation, loop correction and parametric resonance problems eliminates part of the parameter space each, while they three together kill all possible parameter configurations. At least one of those problems therefore has to be fixed to make locked inflation work again. During locked inflation, parametric resonance is disfavored for its exponentially growing energy transfer from $\phi$ field to $\chi$, so it acts as a problem here which may end locked inflation much earlier than expected.

From our above analysis of the mechanism of parametric resonance, we infer for $\phi$ with more than one real component, parametric resonance is inhibited. This solves the parametric resonance problem of locked inflation and makes it possible. Such an argument comes from \refeq{rs-condition}, the condition of rolling stage. For $\phi$ singlet, $\mchi^2$ is dominated by $\lambda\phi^2$ so this condition breaks when $\phi$ is approaching zero. Only when \refeq{rs-condition} breaks, $\phi$ would enter the zero-crossing stage and perform the exponential boost. If $\lambda\phi^2$ isn't the absolute dominant of $\mchi^2$, \refeq{rs-condition} would never break down and the zero-crossing stage would never be entered, even if $\phi=0$. For $\phi$ multiplets, every component of $\phi$ interacts with $\chi$ independently. So when one component crosses zero, other components are in general nonzero and they provide $\chi$ with a large effective mass. The zero-crossing component thus never dominates and \refeq{rs-condition} never breaks.

As an example, we take a $\phi$ doublet, whose components are $\phi_1$, $\phi_2$. For demonstration, we simply assume they have an identical amplitude $|\phi|$ but a phase difference $\frac{\pi}{2}-2\varphi$. So they can be written as
\eqa{\phi_1&=&|\phi|\cos(\mphi t-\varphi),\\
\phi_2&=&|\phi|\sin(\mphi t+\varphi).}
The effective mass squared of $\chi$ then becomes
\eq{\tilde\mchi^2\equiv M^2+2\lambda(\phi_1^2+\phi_2^2)=M^2+2\lambda|\phi|^2(1+\sin2\varphi\sin2\mphi t),}
equivalent to the singlet case with transformed $M^2\rightarrow\tilde M^2=M^2+2\lambda|\phi|^2$, $\lambda\rightarrow\tilde\lambda=\lambda\sin2\varphi$ and $\mphi\rightarrow\tilde\mphi=2\mphi$.

The equivalent singlet case however has the relation $\tilde\lambda|\phi|^2<\tilde M^2$, which breaks the assumption $\lambda|\phi|^2\gg m^2\gg M^2$. In such cases, the condition of rolling stage \refeq{rs-condition} would never break, and the parametric resonance effect would then be inhibited.

In general cases where $|\phi_1|\ne|\phi_2|$ but $|\phi_1|\sim|\phi_2|$,  there are also similar results, which typically have $\tilde M^2\sim0.1\tilde\lambda(|\phi_1|^2+|\phi_2|^2)$. For the same reason, \refeq{rs-condition} would never break and locked inflation can last long enough for $\phi$ multiplets. Therefore for $\phi$ multiplets, the parametric resonance ``problem'' of locked inflation is not actually a problem. Of course, this doesn't solve the problem of embedding locked inflation in some particle physics theory. However, one still needs a viable particle physics setup (or a convincing argument about the decay to visible sector) to make locked inflation promising. This is beyond the coverage of this paper.

\secs{Discussion of Inhomogeneity Amplification}
The cosmic inhomogeneity, if arose from the inflaton, may get amplified by parametric resonance during the large $\delta'$ stage. The amplification also requires a change in the equation of state of the universe that is triggered directly or determined indirectly by the total boost rate of $|\chi|$. This process is similar to modulated perturbations discuss in \cite{Allahverdi:2004bk}. The difference here is the decay rate is determined by the relative phase difference between the inflaton and the preheated field, i.e.\ both fields instead of the inflaton only.

To explicate the whole process, we start with two points at a distance $1/k$. In the beginning of inflation, they live within a Hubble radius, so homogenous $\chi$ is assumed. After the mode $k$ leaves horizon, the two points become uncorrelated and the phases of $\chi$ at the two points start to differ because of the inhomogeneity $\delta\phi_k$ which causes different $\mchi$'s. Also, as we choose the end of inflation at a specific value of $\phi$ on the uniform-density hypersurface, the point with a slightly larger $\phi$ would also gain a slightly longer inflation. For both reasons, at the point where $\phi$ is larger, the phase that $\chi$ experiences would also be larger. This transfers inhomogeneity to $\chi$'s phase from $\phi$'s energy density.

After inflation, the universe enters the large $\delta'$ stage. Due to statistical reasons, the total transformation matrix of the whole large $\delta'$ stage would be an ellipse with the average total boost rate $\langle\acute\eta\rangle$ and the difference between the major and minor axes of the scale $2\Delta\langle\acute\eta\rangle$. Different initial phases of $\chi$ then correspond to different total boost rates, i.e.\ different parametric resonance efficiencies. If there exists a transition of the equation of state depending on the efficiency of parametric resonance, like the backreaction end which changes the universe from matter-like to radiation-like  when $\lambda|\chi|^2\sim m^2$, additional inhomogeneity may be generated.\footnote{During preheating, the universe is matter-like because $\phi$ dominates with its energy density proportional to $a^{-3}$. For preheating ended by backreaction, the dominant potential term becomes $\lambda\phi^2\chi^2$. Therefore $\phi$ and $\chi$ have the same redshift rate. With their redshift rate $a^{-3}$ for constant-mass fields, and their scaling relation with effective mass \refeq{rs-chi-propto-mass}, we can conclude both fields are proportional to $a^{-2}$. This gives the energy density $\lambda|\phi|^2|\chi|^2\propto a^{-4}$ and thus a radiation-like universe. This result is also numerically verified in \cite{Podolsky:2005bw}.} At points where the efficiency is lower for example, more time is needed to boost $\chi$ to the amplitude required for the transition in the equation of state to take place, and subsequently the entrance into a different equation of state is delayed. Therefore there are regions which stay in the larger equation of state longer and suffer a heavier redshift due to universe expansion, and also regions which enjoy a weaker redshift. The difference in redshift rates finally leads to different energy densities at different places, and thus additional curvature perturbation on the uniform-density hypersurface. If the effect is stronger than the existing inhomogeneity, we will find inhomogeneity amplified in this process.

To calculate, we start from inflation in which the inhomogeneity transfers from $\phi$ to the phase of $\chi$. We still adopt the two points at the distance $1/k$ and difference in $\phi$ at the horizon exit of mode $k$ as $\delta\phi_k$. Because we use a universal ending condition for inflation, we align the evolution history of the two points at the end of inflation. Then the only difference $\delta\phi_k$ causes to the phase of $\chi$ is one point enjoys an additional period of inflation as the field $\phi$ rolls from $\phi+\delta\phi_k$ to $\phi$, and this gives an additional phase to $\chi$. Using the slowroll approximation, the resulting phase difference can be calculated
\eq{\Delta\varphi\equiv\sqrt{2\lambda}\,\phi\left|\frac{\delta\phi_k}{\dot\phi_k}\right|=\sqrt\frac{3\lambda}{2\pi}\,\frac{3\mpl H_k^2}{m^3}\,\frac{\delta\phi_k}{\phi_k},\label{eq-ia-dvphi}}
where we use the subscript $k$ to represent values at the horizon exit of mode $k$.

During the large $\delta'$ stage, the parametric resonance process is approximated as stochastic. So we acquire the statistical result of the average total boost rate $\langle\acute\eta\rangle$ and its statistical error $\Delta\langle\acute\eta\rangle$, which are defined in \refeq{db-aeta} and \refeq{db-daeta} and are functions of $n$, the number of half cycles of $\phi$ during the large $\delta'$ stage. If we consider the total transformation matrix of the whole large $\delta'$ stage which transforms a circle of unity to an ellipse, $\langle\acute\eta\rangle$ can also be understood as the typical average total boost rate of the ellipse and $\Delta\langle\acute\eta\rangle$ as half of the typical difference of the total boost rate between the major and minor axes.

Now we have the ellipse as the total boost rate function of the initial phase. The total boost rate goes from maximum to minimum with a change of initial phase $\pi/2$, so the phase difference $\Delta\varphi$ would give a difference in the total boost rate, and thus a difference in the number of half cycles of $\phi$ needed to reach the same amplitude $|\chi|$. For $\Delta\varphi<1$, we have the number of additional half cycles
\eq{\Delta n=\frac{2\Delta\varphi}{\pi}\,\frac{2\Delta\langle\acute\eta\rangle}{\langle\eta\rangle}.}

Then $\Delta n$ leads to delayed or advanced transition of the equation of state. Let us suppose the equation of state is changed by $\Delta w$. The resulting inhomogeneity is then
\eq{\frac{\delta\rho_k}{\rho_k}\equiv\frac{\Delta n\pi}{m}\,3\Delta wH=\sqrt\frac{6\lambda}{\pi}\,\frac{18\Delta w\mpl HH_k^2}{m^4}\,\frac{\Delta\langle\acute\eta\rangle}{\langle\eta\rangle}\,\frac{\delta\phi_k}{\phi_k},\label{eq-ia-drr}}
where $H$ here is the Hubble rate at the transition of the equation of state. 

If we don't want it to overwhelm the existing inhomogeneity $\delta\phi_k/\phi_k$, we then need it to be smaller than that, which gives
\eq{\sqrt\frac{6\lambda}{\pi}\,\frac{18\Delta w\mpl HH_k^2}{m^4}\,\frac{\Delta\langle\acute\eta\rangle}{\langle\eta\rangle}<1,}
and consequently,
\eq{\lambda<\frac{\pi m^8}{1944\Delta w^2\mpl^2H^2H_k^4}\,\frac{\langle\eta\rangle^2}{\Delta\langle\acute\eta\rangle^2}.\label{eq-ia-lambda}}

For $\Delta\varphi>1$, we have instead
\eq{\Delta n=\frac{2\Delta\langle\acute\eta\rangle}{\langle\eta\rangle},}
which generates the inhomogeneity
\eq{\frac{\delta\rho_k}{\rho_k}=\frac{6\pi\Delta wH}{m}\frac{\Delta\langle\acute\eta\rangle}{\langle\eta\rangle}\label{eq-ia-drr2}.}
$\Delta\langle\acute\eta\rangle/\langle\eta\rangle$ is at least 1 and increases as the large $\delta'$ stage continues; $\Delta w\sim\frac{1}{3}$. To make $\delta\rho_k/\rho_k<10^{-5}$, we therefore need at least 8 e-folds before the transition to get a small enough $H$, which is hardly possible for a quadratic slowroll inflation, unless we have thousands of e-folds of inflation to redshift $\chi$. 

So we will still focus on the $\Delta\varphi<1$ case. If we consider the cosmic scale whose $\delta\phi_k/\phi_k\sim10^{-5}$ and $H_k^2\sim10m^2$, we can see from \refeq{ia-dvphi} the constraint from $\Delta\varphi<1$ is already quite strong. For the backreaction end of preheating of the quadratic slowroll inflation, there is $\lambda<10^{-6}$. 

However the constraint from \refeq{ia-lambda} is even stronger. At the backreaction end, the equation of state switches from 0 to $\frac{1}{3}$, so $\Delta w=\frac{1}{3}$. For the preheating fully consisted of the large $\delta'$ stage, we can simplify part of \refeq{ia-lambda} to a function of $\nxx{2}{4}$ only
\eq{\frac{m^2}{H^2}\,\frac{\langle\eta\rangle^2}{\Delta\langle\acute\eta\rangle^2}=\frac{3\sqrt{3\pi}\,\mpl\langle\eta\rangle^2}{4|\phi_2|(\langle\eta^2\rangle-\langle\eta\rangle^2)}e^{\frac{3}{2}\nxx{2}{4}}.\label{eq-ia-part}}
So \refeq{ia-lambda} becomes
\eq{\lambda<\frac{\sqrt{3\pi}m^6}{2592\mpl|\phi_2|H_k^4}\,\frac{\langle\eta\rangle^2}{\langle\eta^2\rangle-\langle\eta\rangle^2}e^{\frac{3}{2}\nxx{2}{4}}.}

Here we simply choose $\nxx{2}{4}=4.5$ and $H_k^2=10m^2$ to get the numerical value $\lambda<10^{-14}$. If we want the result $\lambda<10^{-5}$, from which we can reserve a tiny living space for the backreaction end as shown in \refig{Graph-ReheatingPSpace.pdf}, we would need $\nxx{2}{4}>18$, which means more than $10^{10}$ e-folds of inflation. For the backreaction end with both the large and small $\delta'$ stages during preheating but the same e-folding of the large $\delta'$ stage, the constraint is even stronger.

Such a strong constraint basically rules out the backreaction end of preheating for the quadratic slowroll inflation, unless additional techniques are taken. The attempt of using a dynamic $\lambda$ to solve this problem is also difficult, because that would require $\lambda<10^{-14}$ during inflation, which makes $\chi$ also slowrolling. For natural initial conditions, one then needs to deal with a very large $\chi$ field after inflation which doesn't even need parametric resonance to become large. This however disagrees with our initial purpose of choosing $\chi$ as the first level decay product of $\phi$. Therefore, we infer that the first level of decay of the inflaton of quadratic slowroll inflation is insufficient for reaching a chemical equilibrium between the two types of particles, if we still adopt $\delta\phi/\phi\sim 10^{-5}$ at the cosmic scale.

One may, however, assume inflation only generated a much weaker inhomogeneity and preheating then amplified it to the proper amount. To get a weaker inhomogeneity, we need to change the mass $m$. In order to benefit from the scaling property, we will still use $m_0$ to indicate the old $m$ which alone can get the proper inhomogeneity during inflation. On the r.h.s of \refeq{ia-drr}, $H$ and $H_k$ are both proportional to $m$, and $\delta\phi_k/\phi_k\propto m^2$, so the r.h.s as a whole is proportional to $m$, while the l.h.s is the desired inhomogeneity and should be kept unchanged. We can absorb the dependence on $m$ into $\lambda$ for the rest of the calculation, and drag it out in \refeq{ia-lambda}. If nothing else is modified, we would get a similar result $\lambda m^2/m_0^2\sim10^{-14}$. Still if we want $\lambda\sim10^{-5}$, this would require $m\sim10^{-11}\mpl$. Also, because the amplification rate is the term on the l.h.s of \refeq{ia-drr} before $\delta\phi_k/\phi_k$ which has $H_k^2$ in it, the amplification is scale dependent. Different length scales leave horizon at different Hubble rates which consequently give different inhomogeneity transfer rates. This comes from the different amounts of phase differences of $\chi$, which are only generated at their corresponding horizon exits. Therefore, the scale dependence further adds $-2\epsilon$ to the spectral index for this case, providing $\epsilon\equiv\frac{\dd}{\dd t}\frac{1}{H}$ is the slowroll parameter.

For generic cases, we just need to redefine $\Delta\varphi$. Still aligning every point at the end of inflation and assuming the only difference the inhomogeneity of inflaton gives on $\chi$'s phase is a slightly longer or shorter inflation, we can then define it as $\Delta\varphi\equiv\langle\mchi\rangle\Delta t$. Here $\langle\mchi\rangle$ is the average $\mchi$ during the additional inflationary period $\Delta t$ caused by the inhomogeneity of the inflaton. For short, we use $\delta_k\equiv\delta\rho_k/\rho$ as the energy density inhomogeneity at horizon exit. The time difference of inflation can then be written as
\eq{\Delta t=-\frac{\delta\rho_k\dd t}{\dd\rho}=-\frac{\delta_k\dd t}{\dd\log\rho}=\frac{\delta_k}{2H_k\epsilon_k},}
where again the subscript $k$ indicates the value at horizon exit of mode k. So the inhomogeneity arose from parametric resonance, represented as $\tilde\delta_k$, should be
\eq{\tilde\delta_k=\frac{6\Delta wH}{\mphi}\frac{\Delta\langle\acute\eta\rangle}{\langle\eta\rangle}\frac{\langle\mchi\rangle_k}{H_k\epsilon_k}\delta_k.}
The amplification rate then should be the terms on the r.h.s before $\delta_k$ (for $\Delta\varphi<1$).

Although we have only given one example, this constraint has basically ruled out the backreaction end of parametric resonance in preheating for $m\sim 10^{-6}$. This constraint is also quite strong in general cases, as long as all of the following criteria are met so that the constraint is applicable.
\begin{enumerate}
\item The inflaton has some inhomogeneity after the horizon exit.
\item There is a period of large $\delta'$ stage during the first level decay of the inflaton.
\item A transition of the equation of state of the universe is a direct or indirect consequence of the first level decay product's reaching a critical density.
\end{enumerate}

This constraint however doesn't apply to the small $\delta'$ stage, because all phases converge to the same one in the small $\delta'$ stage. Nor does it apply to perturbative reheating, etc. The constraint doesn't affect models without any transition in the equation of state, such as the inflaton behaves like radiation after inflation which has the same equation of state with the interaction term.

\secs{Summary}
In this article, we have established an analytic framework for the homogeneous mode of parametric resonance in a particular parameter space. In a static background, the Mathieu equation is analytically solved to give the stable and unstable bands. In an expanding universe, distinct stages of large and small $\delta'$ are separately considered. At the large $\delta'$ stage, the phase $\chi$ experiences for every half cycle of $\phi$ is regarded as random and independent of the history, like the Markov chain. During the small $\delta'$ stage, all states with different phases converge to the same eigenstate which is slowly varying. In this eigenstate, $|\chi|$ enjoys a stair-like growth. Both stages share the same average boost rate $\langle\eta\rangle=0.346$. The effect of backreaction is considered, in which a positive feedback acts as the terminator of parametric resonance once it takes over $\mphi$. We also find the backreaction has an impact on $\delta'$ which comes to important earlier than the $\mphi$ take-over and guarantees a short large $\delta'$ stage before the backreaction exit of parametric resonance.

Applying it on preheating, we have constructed a timeline starting from the beginning of inflation till the backreaction exit. Taking the quadratic slowroll model as an example, the large $\delta'$ guarantee point and the transition point from large to small $\delta'$ are included. The violation of $\lambda|\phi|^2\gg m^2$ has also been considered as another exit of parametric resonance, namely the small $|\phi|$ exit. Regarding to how parametric resonance is terminated and whether a small $\delta'$ stage presents, we divide the parameter space of $\lambda$ and $\gamma_0$ into four parts. The backreaction exit typically takes place during the 5th e-fold of preheating for slowroll models with e-folding 60. 

However in a subsequent discussion, we have found parametric resonance during preheating can be responsible in amplifying inhomogeneity. To avoid conflict with current CMB spectrum, it acts as both a constraint on parameter space for some inflationary models, and an amplifier to the weak inhomogeneity of some other models which are otherwise disfavored. We have also set up a demonstration by ruling out the backreaction end of preheating of the quadratic slowroll inflation with mass $10^{-6}$. Amplification rate for general cases is calculated, and criteria of applicability of this effect is explained.

On the other hand, the parametric resonance problem of locked inflation isn't a problem for $\phi$ multiplets. This is because when one component of $\phi$ crosses zero, the contribution from other components on $\mchi$ would be strong and consequently prevent the exponential boost effect.

This paper aims to show an alternative framework for the broad resonance. For this reason, we don't specify the particle physics model and only use singlets. We also show the preheating stage for quadratic slowroll inflation only as an demonstration, without calculating the reheating temperature and thermalization, or discussing difficulties of inflationary models, such as the visible v.s.\ dark sector or how to embed locked inflation in a particle physics foundation.

\acknowledgments{I would like to thank Yeuk-kwan Edna Cheung and Yun Zhang for extensive discussions and Anupam Mazumdar for helpful suggestions on the draft. Exchange of views with Mingzhe Li and Youhua Xu is also acknowledged. This work is supported in part by A Project Funded by the Priority Academic Program Development of Jiangsu Higher Education Institutions (PAPD), NSFC grant No.\ 10775067, Research Links Programme of Swedish Research Council under contract No.\ 348-2008-6049, and Chinese Central Government's 985 grant for Nanjing University.}
\bibliographystyle{jcap}
\bibliography{Main}
\end{document}